\documentclass[%
 reprint,
 amsmath,amssymb,
 aps,
]{revtex4-2}
\usepackage{chemmacros}
    \chemsetup{modules={all}}
\usepackage{bm}
\usepackage{booktabs}
\usepackage{graphicx}
\usepackage{dcolumn}
\DeclareMathOperator*{\argmax}{arg\,max}

\usepackage{bbold}
\usepackage{hyperref}

\renewcommand\vec{\mathbf}

\def\vtheta{{\bm{{\theta}}}}
\def\vthetas{{{\bm{{\theta}}}^\star}}
\def\vxs{{\bf x}_s}
\def\vxe{{\bf x}_e}
\newcommand{\colvec}[2][.8]{%
  \scalebox{#1}{%
    \renewcommand{\arraystretch}{.8}%
    $\begin{bmatrix}#2\end{bmatrix}$%
  }
}
\begin{document}
\def\vtheta{{\bm{{\theta}}}}
\def\vthetas{{{\bm{{\theta}}}^\star}}
\def\vxs{{\bf x}_s}
\def\vxe{{\bf x}_e}

\preprint{APS/123-QED}

\title{Theoretical upper bound of multiplexing in stochastic sensory receptors
}

\author{Asawari Pagare} \thanks{These two authors contributed equally}
\author{Sa Hoon Min} \thanks{These two authors contributed equally}
\author{Zhiyue Lu} \email{zhiyuelu@unc.edu}
\affiliation{%
Department of Chemistry, University of North Carolina-Chapel Hill, NC\\
}%




\date{\today}

\begin{abstract}
Biological sensory receptors provide excellent examples of microscopic scale information transduction amidst stochastic noise. We argue that stochasticity is not always a hindrance to sensing. Instead, it could allow a single stochastic sensor to perform multiplexing: simultaneously transducing multiple types of environmental information to the downstream sensory network. Through a Langevin dynamics simulation of a ligand-receptor sensor in a bath of ligands, we demonstrate that a binary-state receptor can simultaneously encode multiple independent environmental variables, such as ligand concentration and the speed of media flow. We develop a general theory of stochastic sensory multiplexing and suggest two theoretical upper bounds. Furthermore, we conjecture that randomly generated sensors typically saturate the tighter upper bound. The theoretical framework developed in this study, which involves a rank-deficient maximum likelihood analysis (rd-MLE), provides a systematic approach to comprehensively assess a sensor's sensory ability without any initial assumptions. This theoretical framework can inspire the design of more efficient artificial sensors.

\end{abstract}

\maketitle


\section{Introduction}
Sensory receptors are essential to cells. These molecular complexes perceive information from external environments and transmit it into the cell via various signaling mechanisms, despite thermal, and other, noise \cite{antebi2017combinatorial, antebi2017operational, lestas2010fundamental, hinczewski2014cellular}. A classic example is the ligand-receptor mechanism that estimates the ligand concentration. A receptor has the ability to bind with ligands and sense the ligand concentration by estimating the average fraction of them when it is bound with a ligand \cite{berg1977physics,bialek2005physical}. The accuracy of the estimate, as a result, is determined by the ability to average out thermal fluctuations from the desired average binding fraction. 

The accuracy of ligand-receptor sensory mechanisms has been intensively studied at various levels of model complexity. The seminal work of Berg and Purcell \cite{berg1977physics} first discussed the diffusion of ligand molecules around a sensor and showed that there exists a limitation to a receptor's accuracy due to noise. Later, Bialek and Setayeshgar sharpened the Berg and Purcell's accuracy bound by including the receptor's kinetic back actions to the surrounding ligand solution \cite{bialek2005physical}. Kaizu et al. utilized the theory of diffusion-influenced reactions to improve the result \cite{kaizu2014berg}. Recently, Mora and Nemenman developed a theory for temporal concentration sensing where the environment concentration changes over time \cite{mora2019physical}. Efforts to understand ligand-receptor sensing have also been extended to even more complex models involving more than one sensor \cite{carballo2019receptor,hu2010physical}, a single-sensor sensing multiple types of ligands \cite{franccois2019physical, singh2015accurate}, and complex ligand-receptor networks \cite{singh2020universal,singh2017simple,govern2012fundamental}. Additionally, Markovian signal detection \cite{ahuja2020capacity} and receptor diffusion \cite{nguyen2015receptor} have been considered in the studies of ligand-receptor sensory mechanisms.

\begin{figure}
\includegraphics{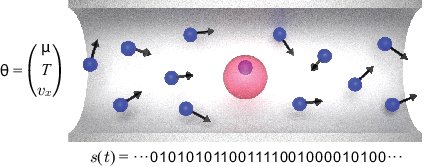}
\caption{A ligand-receptor sensor (big pink sphere) with binary states 0 (unbound state) and 1 (bound with a ligand) immersed in a tube-shaped bath of ligands. The ligand bath is confined in a smooth tube with periodic boundary conditions on its two ends. 
}
\label{fig:system}
\end{figure}

Most previous studies have shared an underlying assumption that the sole sensory task of the receptor is measuring the ligand concentration of the bath $c$, estimated from statistical averaging. For example, sketched in Fig.~\ref{fig:system}, the receptor is immersed in a bath of ligands, let us denote the bound state as $s=1$ and unbound state as $s=0$. Over time, the sensor's dynamics is a binary stochastic sequence $s(t)$. It is straightforward to argue that at the steady state, the local concentration of ligands is approximately equal to the bath's ligand concentration, and the concentration can be estimated by measuring the ratio of the average bound fraction and unbound fraction:
\begin{equation}
    c = \frac{k_+}{k_-} ~e^{-\beta \Delta F} = \frac{\bar s}{1-\bar s} ~e^{-\beta \Delta F} \propto \frac{\bar s}{1-\bar s}
\end{equation}
where $k_+$ is the binding rate constant and $k_-$ is the unbinding rate constant, $\Delta F$ is the free energy change between the bound and the unbound configurations. $\beta=1/(k_B T)$ is the inverse temperature, and we have chosen units such that the Boltzmann constant $k_B=1$. Here the average binding fraction $\bar s$ can be obtained by statistical averaging over a long period of time:
\begin{equation}
     \bar{s}= \lim_{N \rightarrow \infty}\frac{\sum_{i=1}^{N}s(t_{i})}{N}
     \label{eq:bars_stat}
 \end{equation}
In this concentration-sensing regime, one can construct a simple response curve
\begin{equation}
\label{eq:response}
    \bar s = f(c) = \frac{c}{e^{-\beta \Delta F} + c}
\end{equation} that allows one to infer ligand concentration $c$ from the statistical estimation of $\bar s$.

In this work, we recognize that a sensor in realistic environments can be impacted by many facets of the environment in addition to ligand concentration. In Fig.~\ref{fig:system}, a realistic bath (sketched as a tube) is specified by multi-dimensional environmental information $\vtheta = (\mu,T,v_x)$, where the dimensionless $\mu=\ln(c/c^\standardstate)$ is the natural logarithm of the ratio between concentration $c$ and a reference concentration $c^\standardstate$, $T$ is temperature, and $v_x$ is the speed of media flow along $x$-direction. In this regime, the simple statistical estimation of $\bar s$ could be impacted by all environmental factors, and Eq.~\ref{eq:response} becomes
\begin{equation}
    \bar s = g(c,T,v_x)
\end{equation}
Thus, one cannot determine $c$ unless (1) $T$ and $v_x$ are fixed and known, or (2) the response function $g(c,T,v_x)$ is insensitive to $T$ and $v_x$. For the second case, the sensor needs to be designed such that $\partial g / \partial T= 0$ and $\partial g / \partial v_x= 0$. In biology, it has been shown that various processes can behave independently of certain environmental properties \cite{marder2011variability}. An example of such a mechanism is the temperature compensation in circadian clocks \cite{hatakeyama2012generic,kurosawa2005temperature}.

A more ambitious approach, in contrast to the ``variable compensation" mechanism, is to infer all multi-dimensional environmental information from the single sensor. This concept is called multiplexing in information science \cite{thomas2006elements}. \emph{Multiplexing sensing} refers to a single information channel that can transduce multiple independent information channels. In biological sensory networks, it has been recently discovered by Minas et al. that NF-$\kappa$B regulatory networks are capable of performing multiplexing \cite{minas2020multiplexing}. In addition, Singh and Nemenman have demonstrated that a single receptor with two types of outputs can simultaneously report concentrations of two types of ligands \cite{ singh2015accurate}. However, still unanswered is the question that if a simple binary-state receptor could perform general multiplexing, i.e., sensing not only concentrations but also other physical quantities such as temperature or flow speed. 

In this work, we demonstrate that a single receptor can function as a multiplexing information channel and embed different physical properties of the surroundings in its stochastic dynamics. Due to the interaction between the sensor, the local surrounding of the sensor, and the background bath, one can resolve more than one physical quantity from temperature, flow speed, and ligand concentration. Moreover, we develop a general theoretical upper bound of multiplexing and connect it with maximum likelihood estimations \cite{endres2009maximum, minas2020multiplexing, Cover2012-zz} that are rank-deficient \cite{khamaru2019computation, robertson2007maximum}.

The paper is ordered as follows. First, we demonstrate via an in silico experiment that a simple binary-state receptor can simultaneously encode multiple physical quantities, such as ligand concentration and flow speed. Then, we provide a universal theory of multiplexing through which we predict a theoretical upper bound for multiplexing. We also argue that the upper bound can be sharpened if the number of symmetries in the system is known. Then, we develop a rank-deficient maximum likelihood estimation (rd-MLE) approach to comprehensively estimate the sensing capability of an arbitrary stochastic sensor.

\section{Multi-channel Information Inference by Ligand-Receptor Sensor}
\label{sec:2}
In this section, we utilize a simple example of a ligand-receptor sensor to demonstrate that even if a simple sensor does not compensate for changes in irrelevant parameters, it is still possible that the sensor can perform multiplexing sensing and infer multiple environmental variables.

\subsection{In silico experiment of ligand-receptor sensor.}
Here we demonstrate an in silico experiment of an individual ligand-receptor sensor within a ligand bath (see Fig.~\ref{fig:system}). The system is characterized by ligand concentration $c$, temperature $T$, and flow speed $v_x$. The bath tube has a periodic boundary condition at the two ends allowing for a steady media flow along the $x$-axis. In the simulation, we control the flow speed along the $x$-axis direction ${\bf n}_x$ with magnitude $v_x$, while keeping the receptor fixed at the center. The free ligand particles within the bath are simulated by over-damped Langevin dynamics that capture the stochastic Brownian motions, deterministic drift due to $v_x$, and interaction with the receptor at the center:
\begin{equation}
\label{eq:Langevin}
    \dot{\vec{r}_i}= v_x {\bf n}_x+ \frac{1}{{m}\gamma}\vec{F}_i+\vec{\eta}(t)
\end{equation}
where $\vec{r}_i$ is the position of the $i$-th ligand molecule, $m$ is its mass, $\gamma$ is the friction coefficient, $v_x {\bf n}_x$ is the velocity of media along the $x$-axis unit vector ${\bf n}_x$, $\vec{F}_i$ is the total deterministic force experienced by each ligand molecule (in a detailed simulation, this force accounts for the sum of pairwise interactions between ligand pairs and the ligand-receptor attraction), and $\vec{\eta}$ is the Gaussian noise characterized by $\langle\eta_i(t)\eta_j(t')\rangle = 2\gamma k_BT\delta_{ij}\delta(t-t')$, where $k_B$ is the Boltzmann constant (See Appendix~\ref{Sec:AA} for details of the simulation). The ligand--ligand interaction can adopt a short-range repulsive WCA potential \cite{weeks1971perturbation} or can be modeled as free Brownian particles to reduce the computational cost, this simplification does not qualitatively change the result (as shown by Figs.~\ref{fig:WCA_T_c},\ref{fig:WCA_v_c},\ref{fig:WCA_T_vx} in Appendix~\ref{Sec:AB}).

We assume that the receptor can bind with up to one ligand molecule. The binding and unbinding events are modeled by stochastic dynamics following two Poisson rates:
\begin{equation}
k_+= n_l e^{-\beta (E_b-E_{\rm off})}
\end{equation}
where $n_l$ is the number of ligands within a cutoff distance $r_c$ from the receptor, $\beta=1/(k_BT)$ is the inverse temperature, $E_{\rm off}$ is the energy of an empty receptor, and $E_b$ is the energy barrier for the binding-unbinding transition,

Notice that $n_l$ is a fluctuating dimensionless quantity related to a fluctuating local concentration $c'$ as
\begin{equation}
\label{eq:nl}
    n_l=\frac{4\pi r_c^3}{3 V_{\text{tube}}}\frac{c'}{c^\standardstate}
\end{equation}
where $V_{\text{tube}}$ is the volume of the whole simulation tube, $r_c$ is the cutoff radius around the receptor, and we have chosen a reference concentration (unit concentration) $c^\standardstate$ defined by the situation where there is only one ligand molecule within the whole tube. To estimate binding rate $k_+$ in the Langevin dynamics simulation, we do not explicitly evaluate $c'$ but only need to count $n_l$, which is the number of ligand particles within the cutoff radius. (See Appendix~\ref{Sec:AA} for the numerical value of this cutoff radius and the dimensions of the simulation tube.) We can also define the unbinding rate as
\begin{equation}
k_- = e^{-\beta (E_b-E_{\rm on}-\gamma\alpha v_x)}
\label{eq:unbinding_rate}
\end{equation}
where $E_{\rm on}$ is the energy of the receptor when it is bound with the ligand. In the numerical simulation, $c'$ reflects the local concentration experienced by the receptor, which may not be equal to the background ligand concentration $c$ (see Fig.~\ref{fig:system-velocity}).

To describe the media flow's impact on the unbinding event, we assumed that the unbinding rate is boosted by the frictional drag that the bound ligand experiences as it leaves the receptor. The drag force is modeled by $\gamma v_x$. When an unbinding event occurs, the bound ligand displaces from the center of the receptor by a displacement of $\alpha$ in the downstream direction before it is considered as unbound. Thus an approximated dissipative work of $\gamma \alpha v_x$ assists the unbinding event. When the flow's impacts on the binding and unbinding rates are negligible, one can simply remove the $\gamma \alpha v_x$ term, and the system satisfies the detailed balance condition. Notice that by ignoring the dissipative work, the flow speed does not explicitly impact the sensor's dynamics. However, without an explicit dependence on the speed, the sensor can still perform multiplexing to sense both concentration and flow speed (see Fig.~\ref{fig:nodrag} in Appendix~\ref{Sec:AC}).

\subsection{Simple ligand-receptor sensors do not compensate for $T$ and $v_x$.}
The response function for a ligand-receptor sensor $\bar s$ should, in general, depend on all environmental variables, i.e., $c$, $T$, and $v_x$. In other words, a sensor's ability to respond to $c$ can not compensate for changes in environmental condition $T$ and $v_x$.

On the one hand, in a thermally equilibrated environment without media flow ($v_x=0$), it is clearly shown that the response function Eq.~\ref{eq:response} depends both on $c$ and $T$. On the other hand, at a constant temperature environment, the average binding fraction $\bar s$ would depend on the flow speed $v_x$. Such dependence can only disappear by removing the dissipative work term in the unbinding rate Eq.~\ref{eq:unbinding_rate}. Such flow-speed dependence can be seen in Fig.~\ref{fig:cross}a. 
In summary, knowing the single response function $\bar s$ is insufficient for simultaneously determining the values of both $c=c^\standardstate e^\mu$ and $v_x$ (or $T$).

\subsection{Simultaneously inferring multiple parameters (multiplexing).}

Here we demonstrate that by extracting the sensor's trajectory information, one can simultaneously infer multiple environmental information (e.g., $c$, $T$, and $v_x$), and there is no need for compensation. 

In addition to the existing response function $\bar s$, one can extract from the sensor's trajectory $s(t)$, an additional response function:  
\begin{equation}
     C_1(\mu,T,v_x) = \frac{\langle s(t_i)s(t_i+1)\rangle}{\langle s(t_i)s(t_i)\rangle}
\end{equation}
where $C_1$ is the auto-correlation of the sensor's state $s(t)$ evaluated at a time lag equal to unity. A downstream reaction network can realize such a response function with a delayed-on activation: the downstream signaling reaction is only turned on if the sensor remains at the bound state for times longer than a time lag. Such a delay-activated signaling network is indeed less common than that of the simple response $\bar s$, it can be realized by kinetic proofreading reactions \cite{hopfield1974kinetic,qian2006reducing,murugan2012speed}. 
As it will become obvious in the next subsection, the specific choice of the time lag in $C_1$ (i.e. the delay time set by the kinetic-proofreading-like reaction) will only impact the sensitivity of multiplexing but does not qualitatively change the multiplexing behavior.  

\begin{figure}
\includegraphics{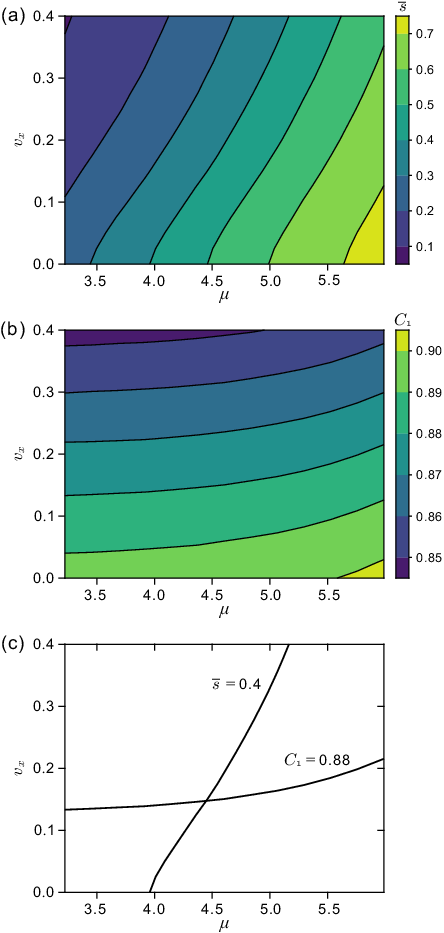}
\caption {By sampling 2-point trajectories at simulations performed at various environmental conditions, one obtains the contour plot of $\bar s(\mu,v_x)$ shown in (a), and of $C_1 (\mu,v_x)$ shown in (b). If one obtains the values of $\bar s$ and $C_1$ from a set of measured trajectories at an unknown environmental condition, then $(\mu,v_x)$ can be simultaneously inferred by crossing the two contour lines shown in (c).}
\label{fig:cross}
\end{figure}

Multiplexing is demonstrated by a contour-line crossing technique shown in Fig.~\ref{fig:cross}c.  When both response functions $\bar s$ and $C_1$ are obtained from the statistics of the sensor's trajectory $s(t)$, one can simultaneously determine both flow speed $v_x$ and ligand concentration of the bath $c={c^\standardstate } e^\mu$ by contour-line crossing (see Fig.~\ref{fig:cross}). Notice that even when the flow speed does not impact the binding and unbinding rates, the contour-line crossing still demonstrates that the sensor can simultaneously infer both concentration $c={c^\standardstate } e^\mu$ and flow speed $v_x$ (see Fig.~\ref{fig:nodrag} in Appendix~\ref{Sec:AC}). 
Similarly, with the two response functions, a receptor can do multiplexing for not only $c$ and $v_x$, but also $c$ and $T$ or $T$ and $v_x$ (see Appendix~\ref{Sec:AD}). A universal theory of multiplexing and its upper bound is discussed in Sec.~\ref{sec:3}

\subsection{Kinetic intuition behind inexplicit speed sensing}
\begin{figure}
\includegraphics{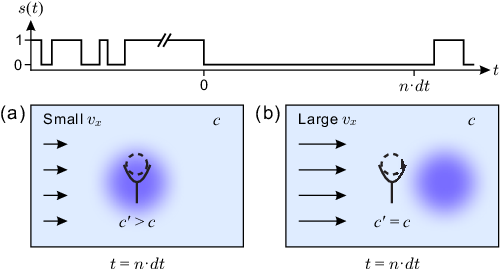}
\caption{Effect of flow speed on the local concentration $c'$ (Eq.~\ref{eq:nl}) of ligands experienced by a receptor. (a) In the small speed regime, the local concentration evolution after a ligand is returned to the bath via receptor-unbinding is dominated by diffusion, shown as the blue Gaussian cloud. (b) In the large speed regime, the local concentration of ligands around the receptor is rapidly replenished by the background bath, and the added ligand from the unbinding event is flushed away from the receptor.}
\label{fig:system-velocity}
\end{figure}

It may appear surprising that flow speed $v_x$ can be sensed even if the kinetics of the sensor does not depend on the flow speed (Appendix~\ref{Sec:AC}).
Here we illustrate in Fig.~\ref{fig:system-velocity} that the flow speed $v_x$ impacts the relaxation of the local ligand concentration $c'$ around the sensor and thus can be inferred by the time-correlation of the receptor's dynamics. 
Consider that the local concentration $c'$ is initially equal to the background ligand concentration $c$. Right after an unbinding event ($s$ transitioning from 1 to 0), one additional ligand is deposited back into the local bath, causing a temporary increase of local concentration $c'>c$. Then over time, the local concentration $c'$ tends to relax back to $c$ as shown by the dispersing blue blob in Fig.~\ref{fig:system-velocity}. The rates of such relaxations depend on both the flow speed and the diffusion of the ligands.
If the velocity is close to zero, as in Fig.~\ref{fig:system-velocity}a, the $c'$ relaxation is dominated by diffusion alone, and thus $c'>c$ for a long period of time. In this time period, the receptor has a higher possibility of binding with a ligand due to $c'>c$. In comparison, if the flow speed is large, as in Fig.~\ref{fig:system-velocity}b, then the additional ligand that unbinds from the receptor quickly moves in the x direction ${\bf n}_x$, and the local concentration $c'$ is replenished by the background medial $c$. As a result, lower flow speed could result in stronger auto-correlation in binding events, and higher flow speed removes this auto-correlation by quickly resetting local concentration $c'$ back to the background concentration $c$.

The analysis above illustrates that even if the flow speed $v_x$ does not impact the unbinding rate (i.e., no dissipative work term in Eq.~\ref{eq:unbinding_rate}), the sensor is still able to perform concentration and flow speed multiplexing by using the proposed contour-line crossing. (See Fig.~\ref{fig:nodrag} in Appendix~\ref{Sec:AC}) 

\section{Universal Theory of Multiplexing by A Stochastic Sensor}
\label{sec:3}
In this section, we theoretically derive a simple yet universal upper bound of multiplexing for generic biological sensors (Eq. ~\ref{eq:uni_up}). 
Furthermore, we provide a systematic approach (rank-deficient maximum likelihood analysis) to analyze a sensor's sensitivity in sensing multiple environmental variables.

\subsection{Theoretical Upper Bound for Multiplexing}
Let us start with a general description of a microscopic sensor sensing information from its environment: a single sensor with internal degrees of freedom $\vxs$ interacts with a stationary stochastic environment with the degrees of freedom denoted by $\vxe$. The environment is characterized by a few stationary macroscopic variables that are denoted by the $n$-dim vector $\vtheta$ (e.g., temperature, chemical concentration, pressure, flow speed, pH, etc.). Notice that even though the macroscopic parameters are fixed, the environment's microscopic states $\vxe$ evolve erratically in time due to thermal fluctuations and the interaction with the sensor. The sensor, through its interaction with the environment, can read environmental information $\vtheta$ and transduce such information to the downstream sensory network. In summary, from a microscopic perspective, both the sensor and environment evolve stochastically due to the thermal fluctuations; the state of the environment can impact the dynamics of the sensor, and the sensor's state change can in turn perform a \emph{back action} to the local environment (see Fig.~\ref{fig:system-velocity}). A general equation of motion of the composite system of the sensor and the environment's microscopic state can be written as
\begin{equation}
\label{eq:full_eom}
    \frac{{\rm d} (\vxs,\vxe)}{{\rm d} t } = {\bf F}(\vxs,\vxe;\vtheta) 
\end{equation}
where ${\bf F(\cdot;\vtheta)}$ is a function of both $\vxs$ and $\vxe$ and is parameterized by the stationary environmental properties 
$\vtheta=(\theta_1,\cdots,\theta_n)$ such as temperature, concentration, etc.
Here, $n$ denotes the number of independent environment variables. For biological sensing, the composite system of a sensor and environment is in contact with a larger background thermal environment, and thus one can interpret the above equation as Langevin dynamics where ${\bf F}(\vxs,\vxe;\vtheta)$ contains both the deterministic dynamics and the stochastic terms due to the thermal fluctuations. Notice that not all information in the microscopic state of the sensor (i.e., the full solution to Eq.~\ref{eq:full_eom}, $\vxs(t)$) can be transduced to the downstream network. The limitations to accurately sensing $\vxs$ at infinitely high time resolution can be formulated by the following two assumptions:

{\bf{Discrete-state sensor assumption}}: $\vxs(t)$ represents the full microscopic state of the sensor at time $t$ (i.e., the locations and momentum of each atom in the receptor molecule), and it is unrealistic to expect the downstream bio-sensory network to observe or record a full micro-state trajectory of the sensor. Rather, the downstream sensory network may only read the sensor as in one of a few coarse-grained states $s(\vxs)$. As a consequence, the coarse-grained state $s(\vxs)$ reduces the high-dimension micro-state space of $\vxs$ into a discrete state space. For a simple receptor, one can naturally assume that the sensor's coarse-grained state is binary: $s\in \{0,1\}$, which assumes state `0' when the receptor is unbound from any ligand, and assumes state `1' when the receptor is bound to a ligand. Given the thermal fluctuations and the stochastic dynamics of $\vxs(t)$, we take an assumption that $s(t)$ is a discrete-state stochastic process.

{\bf{Discrete-time trajectory approximation}}: In practice, accurately recording a continuous-time trajectory $s(t)$ with infinitely high time resolution comes with infinitely high information cost and thermodynamic cost, and is unrealistic for living organisms \cite{horowitz2014second}. Rather, typical downstream sensory networks can be considered as a noisy ``kernel'' \cite{hinczewski2014cellular} with a finite time-resolution and finite memory lifespan, applied to the signal $s(t)$ over a long time. The output of the kernel integral is the accumulated downstream signaling molecules generated by the downstream network, whose rates depend on the sensor state $s(t)$. On the one hand, consider a kernel whose memory lifespan is very short compared to the transition time of the sensor; then, even if it has infinitely high time resolution, the accumulated output from the ``kernel'' is no more than single-time-point statistics. In this case, information about the time-auto-correlation information of $s(t)$ is lost. On the other hand, if the kernel has a very long memory lifespan but an extremely low time resolution (i.e., sampling time lag is approximately equal to the memory lifespan), then effectively, the kernel can at most encode information of two-time trajectories with time lag equal to the memory lifespan. With these considerations, we can simplify the downstream reaction network, given its time-resolution and memory lifespan, into a statistical estimator based on $n_t$-point discrete-time trajectories. In other words, only ``visible'' to the downstream network is the accumulated statistics of an ensemble of discrete-time trajectories $S$'s, where each $S=\left( s(1),s(2),\cdots,s(n_t) \right)$ is a discrete-time trajectory, whose trajectory time length $n_t$ is limited by the kernel's memory length divided by the kernel's time resolution.\footnote{ Notice that this argument is general and the time-lag between $s(i)$ and $s(i+1)$ does not need to be equal to lag between $s(j)$ and $s(j+1)$, for $1\leq i,j \leq n_t-1$. The effective time lags and effective time points $n_t$ are limited by the self-correlation of the trajectory $s(t)$ and by the property of the downstream kernel.} 

\begin{figure*}
\includegraphics{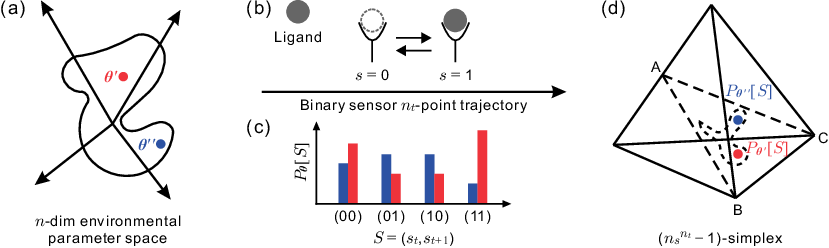}
\caption{A complex environment can be characterized by $n$ independent variables $\bm\theta$, forming a $n$-dim parameter space shown by (a). At a given condition $\bm\theta$, the ligand-receptor sensor sketched in (b) evolves with a stochastic trajectory $s(t)$. From the statistics of 2-point trajectories $S= ``00", ~``01",~ ``10",~\text{or } ``11"$ one can obtain the 2-point trajectory probabilities $P_{\bm\theta}[S]$, shown in (c) for two choices of $\bm\theta$'s. As shown in (d), when $n_s=2$ and $n_t=2$, the trajectory probability lives in a 3-simplex. Moreover, due to $P_{\bm\theta}[01]=P_{\bm\theta}[10]$ (see Appendix~\ref{Sec:AE}), the probability space is reduced to a 2-dim triangle ABC within the 3-simplex. The sensor then establishes a map between the $\vtheta$ and probability spaces: an ensemble of environmental conditions $\vtheta$'s (shown by the solid blob in (\textbf{a})) is mapped to an ensemble of $P_{\bm\theta}[S]$'s (shown by a dashed blob in the 2-dim triangle ABC in (\textbf{d})). In this example, $n_s=2$ and $n_t=2$, because the probability has an irreducible dimension 2, the sensor can infer up to 2 independent environmental variables.  }
\label{hilbert}
\end{figure*}

Under the ``Discrete-State Sensor Assumption'' and the ``Discrete-Time Trajectory Approximation'', the information visible to the downstream sensory network over a long period of time is a collection of discrete trajectories $\{S\}$, where $S=\left( s(1),s(2),\cdots,s(n_t) \right)$. 
For a sensor with $n_s$ coarse-grained states and trajectory time length of $n_t$, there are $n_s^{n_t}$ different possible trajectories, e.g., for $n_t=2$, a binary state sensor ($n_s$=2) can produce $S\in\{00,~01,~10,~11\}$.
Via the statistics over all possible trajectories, the sensor maps information of the $n$-dim environmental parameter into a trajectory probability simplex, and if the dimension of the probability simplex is lower than $n$, not all environmental information can be fully determined. 

{\bf Universal upper bound of multiplexing.} The statistical map of the sensor allows us to obtain a universal upper bound of multiplexing for any arbitrary sensor-environment system. With the statistics of the trajectories, a sensor establishes a map from the $\vtheta$-space into the trajectory probability space of $P_{\bm\theta}[S]$ confined in a $d_x(n_t,n_s)\equiv (n_s^{n_t}-1)$ dimensional simplex. As illustrated by Fig.~\ref{hilbert}, a $n$-dim solid blob in $\vtheta$-space (an ensemble of environments) is mapped by a sensor to the dashed blob in the trajectory probability space. The sensor's ability to infer $\vtheta$ is thus limited by one's ability to construct an inverse map from $P_\vtheta[S]$-space back to the $\vtheta$-space. 
Furthermore, we assume that the microscopic sensor is well below the thermodynamic limit and thus free from any discontinuity (e.g., caused by phase transition \cite{arndt2000yang}). To infer $\vtheta$ from the trajectory probability, one needs to find an injective continuous inverse map, which can only be constructed from the $d_x(n_t,n_s)$-dim simplex to a sub-$\vtheta$-space of dimension lower than or equal to the simplex's dimension. As a result, the maximum independent environmental degrees of freedom that one can infer from the sensor must obey a \emph{theoretical upper bound of multiplexing}:
\begin{equation}
    \label{eq:uni_up}
    r_s \leq d_x(n_t,n_s) \equiv n_s^{n_t}-1
\end{equation}
Such upper bound predicts the maximum number of independent environmental variables that one can infer from a sensor's statistics of $n_t$-time trajectories $\{S\}$. 

\subsection{Tightened upper bound of multiplexing.}
The universal upper bound, Eq.~\ref{eq:uni_up}, can be sharpened if more information of the sensor-environmental system is known. For specific sensors with a given number of coarse-grained states and given connectivity in their state space, the trajectory probabilities may be further restricted by symmetries. The number of independent symmetries $d_{sy}$ reduces the trajectory probability space from the full $d_x(n_t,n_s)$-dim simplex to a lower-dimension subspace of irreducible dimension $d=d_x-d_{sy}$. Based on the continuity argument, we can sharpen the upper bound of multiplexing as
\begin{equation}
\label{eq:tight}
    r_s \leq d_{ir}(n_t,n_s) \equiv n_s^{n_t}-1 - d_{sy}
\end{equation}

Notice that the number of symmetries $d_{sy}$ can not be expressed in a closed analytical form for arbitrary sensors as it can depend on the topology of the sensor-environment kinetics. However, for simple two-state sensors, we can find the value of the $d_{sy}=1$ for $n_t=2$ and $n_t=3$:
In the example of $n_t=2$ and $n_s=2$ all possible trajectories are $``00",~``01",~``10",$ and $``11"$ and we can easily argue the symmetry of $P_\vtheta[01]=P_\vtheta[10]$: Consider that we obtain the statistic probabilities of trajectories by chopping an infinitely long trajectory into many two-time segments. The number of $``01"$ segments and the number of $``10"$ must be equal or differ by $\pm 1$ (also discussed in Appendix~\ref{Sec:AE}). As a result, given sufficient statistics, the frequency of $``01"$ must be equal to that of $``10"$.  
Given the above analysis, we can obtain the general sharper bound of multiplexing by two-state sensors reporting 2-point trajectories as
\begin{equation}
    r_s(n_s=2,n_t=2) \leq 2^2-1 -d_{sy}(n_t=2,n_s=2) =2
\end{equation}
where $d_{sy}(n_t=2,n_s=2)=1$. This agrees with our observation in the ligand-receptor system: from 2-point trajectories of the receptor, one can simultaneously infer only 2 out of 3 environmental variables. In this case, shown in Fig.~\ref{hilbert}(d), the irreducible trajectory probability space is a 2-dimensional plane (ABC) in the 3-dimensional simplex. 

For $n_t=3$ and $n_s=2$, the $7$-simplex is reduced to a $4$-dim manifold due to $d_{sy}=3$ symmetries, similar to the argument for the 2-point trajectories, one can find that there are 3 symmetries:  $P_{\vtheta}[001]=P_{\vtheta}[100]$, $P_{\vtheta}[011]=P_{\vtheta}[110]$, and $P_{\vtheta}[001]+P_{\vtheta}[101]=P_{\vtheta}[010]+P_{\vtheta}[110]$ (see Appendix~\ref{Sec:AE}).
In this case, for binary state sensors:
\begin{equation}
    r_s(n_s=2,n_t=3) \leq 2^3-1 -d_{sy}(n_t=3,n_s=2) =4
    \label{eq:3point}
\end{equation}
where $d_{sy}=3$. One can argue that if 3-point trajectory statistics are obtained from a sensor, it could be used to infer up to 4 independent environmental variables. 

In the Discussion, it is conjectured and demonstrated, through a simple Markov model, that a simple sensor can saturate the tightened upper bound with a probability of approximately 1. 

\subsection{Rank-deficient maximum likelihood analysis}

The contour-line crossing technique in Sec.~\ref{sec:2} demonstrates that the ligand-receptor sensor can perform velocity-concentration multiplexing. However, it is important to develop a general framework to determine (1) how many independent environmental variables the receptor can simultaneously sense, and (2) how accurately can each environmental property be sensed. This can be done with the rank-deficient maximum likelihood analysis of two-step trajectories.

The choices of the two response functions $\bar s$ and $C_1$ in the previous section rely on the physical intuition of a specific system. They thus can not be easily generalized to other types of sensors or other tasks. To provide a general framework, here we utilize trajectory probability for environment $\vtheta$, $P_{\vtheta}[S]$ as a systematic representation of the response functions to study multiplexing in general. The trajectory $S$ represents the sensor-state trajectory $S=s(t_0),~s(t_1),~s(t_2),\cdots$. It can be shown that the response functions used in the last section can both be obtained from trajectory probabilities of 2-point trajectories of time lag equal to unity:
\begin{equation}
    \bar s (\vtheta)= \frac{P_\vtheta[01]+P_\vtheta[10]}{2}+P_\vtheta[11] 
    \label{eq:sbar-trj}
\end{equation}
and
\begin{equation}
    C_1 (\vtheta) = \frac{P_\vtheta [11]}{(P_\vtheta[01]+P_\vtheta[10])/2+P_\vtheta[11]}
    \label{eq:C1-trj}
\end{equation}

Given the trajectory probability distributions and the ability to perform statistical measurements of various sensor-state trajectories, one can adopt a non-conventional maximum likelihood estimation (MLE) analysis to characterize the number of independent variables that can be sensed by the sensor as well as the estimation accuracy. The conventional MLE method has been proven very useful in estimating the sensory capability of biological sensors \cite{endres2009maximum, minas2020multiplexing}.

Here we utilize a rank-deficient MLE (rd-MLE) analysis to characterize the sensor's information capability. Consider a fixed multi-dimensional environmental condition $\vtheta^*$, let the sensor observe and accumulate $m$ independent trajectories of the sensor's state, $\{S^{*1},S^{*2},\cdots S^{*m}\}$, without knowing the values of $\vtheta^*$. One can perform an MLE of $\vtheta^*$ where the log-likelihood function (llf) of environmental variables $\vtheta$ is defined as
\begin{equation}
l(\vtheta)\equiv\frac{1}{m}\sum_{i=1}^{m}\ln P_{\vtheta}[S^{*i}]
\end{equation}
where the probability of observing each possible trajectory is denoted by $P_{\vtheta^*}[S]$. 
In the limit of abundant sampling, $m\rightarrow \infty$, the llf reduces to a simple expression depending on the trajectory probabilities
\begin{equation}
\label{likep}
l(\vtheta) = \sum_{S} P_{\vthetas}[S]\ln P_{\vtheta}[S]
\end{equation}
The conventional MLE theory then states that the accurate estimation of the environmental variables are achieved at the maximum of the log-likelihood function: 
\begin{equation}
    \bm{\vtheta}_{{\rm est}} = \argmax\big(l(\vtheta)\big)
\end{equation}

In this work, we do not assume the existence of a solution to the optimization problem defined in the traditional MLE, where the Hessian matrix of $l(\vtheta)$ is negative-definite. Instead, we acknowledge that the sensor may not be able to infer all degrees of freedom of the unknown parameters, as $l(\vtheta)$ may not be a strictly concave surface, and the Hessian may not be full-rank. Here we argue that the number of independent degrees of freedom that the sensor can infer from the repeated statistics of trajectories is determined by the rank of the Hessian matrix, $r_s$. Here $r_s$ equals the number of non-zero (negative eigenvalues) of the Hessian matrix.

Here we demonstrate that the non-zero-eigenvalues and their corresponding eigenvectors contain information on the sensor's sensitivity to different types of environmental variables: the magnitude of the negative eigenvalues corresponds to the sensory accuracy and thus the sampling size necessary to determine the environmental status along the direction of the corresponding eigenvector. The eigenvector's direction corresponding to the largest magnitude eigenvalue reflects the most sensitive variable that the sensor can sense. This is demonstrated through numerical calculations on the data obtained through simulation in the following section.

\section{Discussion}

\subsection{rd-MLE analysis on Ligand-receptor sensor}

Recall the Eqs.~(\ref{eq:sbar-trj}) and (\ref{eq:C1-trj}), measuring $\bar s$ and $C_1$ is equivalent to sampling the trajectory probabilities of 2-point trajectories $S=$ ``00'', ``01'', ``10'', and ``11'' where the time lag between the two time points is unity (which is 100 time-steps in the numerical simulation with time resolution $dt=0.01$). Using the statistics of 2-point trajectories, we numerically obtain the llf function and the Hessian matrix for environmental condition $(\mu,T,v_x)$. As predicted by the tightened upper bound, the 3-by-3 Hessian matrix is rank-deficient with rank $r_s=2$, indicating that the statistics of 2-point trajectories can only infer up to two out of three independent environmental variables among $(\mu,T,v_x)$.

To demonstrate the rank-deficient MLE, we numerically obtained \footnote{Derivatives in the $\vtheta$-space are evaluated as finite differentials with increments $\delta\theta = (4.8459\%\mu^*, 5\%T^*, 5\%v_x^*)$. The 2-point trajectory probabilities used in the analysis are obtained by chopping a long simulation $10^{10}$ steps (each step increases time by $dt=0.01$). Each 2-point trajectory is sampled from two time points with the time difference of $100$ steps which is equivalent to time lag $100 dt=1$.} the Hessian matrix of the llf $l(\vtheta)$ at a given condition ${\vtheta}^{*}=(\mu^*,T^*,v_x^*)=(\ln{100},1.0,0.2)$ 
(See Appendix~\ref{Sec:AF} for numerical details of the simulation):
\begin{equation}
\label{eq:hessian}
\colvec[.7]{(-2.24\textbf{7} \pm0.003)\times10^{-1} & (-3.\textbf{6}\pm0.1)\times10^{-2} & (6.9\textbf{5}\pm0.02)\times10^{-1}\\(-3.\textbf{6}\pm0.1)\times10^{-2} & (-2.25\textbf{4}\pm0.007)\times10^{-1} & (-2.4\textbf{7}\pm0.02)\times10^{-1}\\(6.9\textbf{5}\pm0.02)\times10^{-1} & (-2.4\textbf{7}\pm0.02)\times10^{-1} & (-2.2\textbf{3}\pm0.03)}
\end{equation}

We numerically verify that the Hessian matrix is rank-deficient, with two negative eigenvalues ($\lambda_1$ and $\lambda_2$) and a zero eigenvalue ($\lambda_3$):
\begin{align}
    \lambda_1^{\mu, T,v_x}&=-2.4\textbf{8}\pm0.03\\
    \lambda_2^{\mu, T,v_x}&=(-2.06\textbf{0}\pm0.008) \times 10^{-1}\\
    \lambda_3^{\mu, T,v_x}&=(\textbf{2}\pm4)\times 10^{-3} \approx 0
\end{align}
The above Hessian is rank $r_s=2$ and negative-semi-definite, and thus only up to 2 variables can be inferred. 

The above analysis demonstrates that by measuring statistics of 2-point trajectories, a single receptor can simultaneously report two out of three environmental variables, which is in agreement with the predicted tightened upper bound of multiplexing for a binary sensor reporting two-point trajectories.  We can apply the analysis to the receptor by obtaining the eigenvectors corresponding to the 3 eigenvalues starting from the smallest eigenvalue to the zero eigenvalue:
\begin{widetext}
\begin{align} 
\vec{v}_1^{T} &= \{(-2.9\textbf{5}\pm0.04)\times10^{-1}, (1.0\textbf{9}\pm0.02)\times10^{-1}, (9.4\textbf{9}\pm0.01)\times10^{-1}\}\\
\vec{v}_2^{T} &= \{(2.1\textbf{5}\pm0.04)\times10^{-1}, (9.75\textbf{5}\pm0.008)\times10^{-1}, (-4.\textbf{5}\pm0.2)\times10^{-2}\} \\
\vec{v}_3^{T} &= \{(9.3\textbf{1}\pm0.01)\times10^{-1},(-1.9\textbf{1}\pm0.04)\times10^{-1},(3.1\textbf{1}\pm0.04)\times10^{-1}\} 
\end{align}
 \end{widetext}

Given the limitation of two-time trajectories and choice of time lag unity, the dominant eigenvector $\vec{v}_1$ is in most alignment in the direction $(0,0,1)$, which corresponds to the flow speed $v_x$, and then the direction corresponding to $\mu$; the second dominant eigenvector is in most alignment with $T$ and then $\mu$. In other words, if the downstream signaling network receives information of 2-point trajectories rather than a simple time-average $\bar s$, then a ligand-receptor sensory system is most sensitive to flow speed $v_x$, and only secondary sensitive to the logarithm of the concentration, $\mu$.

This general framework or rd-MLE and the procedures demonstrated above can be applied to other stochastic receptors or sensors. It reveals the number of independent variables that a sensor can sense; it also numerically estimates the sensitivity of each variable. Moreover, by repeating the procedures for different trajectory length ($n_t$), time lag, and even the number of sensor states that is distinguishable to the downstream network ($n_s$), one can optimize the downstream reaction network to better infer a desired environmental variable. In the following, this approach is applied to a general Markov model of a sensor-environment system.

\subsection{Conjecture of upper bound saturation}

{\bf General Markov model.} Let us study a generic Markov model of a sensor--environment system. Within the Markov model framework, we focus on randomly generated sensors and analyze their multiplexing performance by using the rd-MLE approach described in this paper. 

\begin{figure}[h!]
\centering
\includegraphics{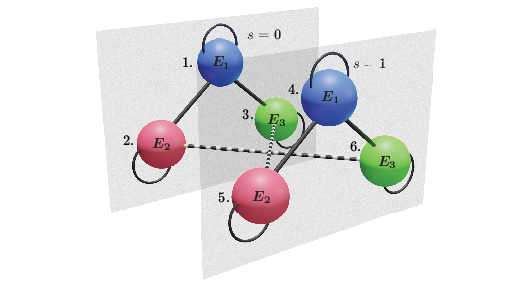}
\caption {A simple 6-state Markov model constructed by combining a 3-state environment with a binary-state sensor. Each transparent plane denotes one state of the sensor ($s=0$ or $1$) and each sphere on a plane denotes one possible environmental state $E_1$, $E_2$, and $E_3$. The transition between $E_1$ and $E_2$, and between $E_1$ and $E_3$ (two solid lines) correspond to the dynamics of the environment independent of the sensor's state. The dashed lines denote the transition between environmental states $E_2$ and $E_3$, which is accompanied by the state change of the sensor. After each time step, the system could remain unchanged, which is characterized by curved loops starting and ending at the same sphere. The transition probability for each allowed transition is summarized into a transition probability matrix in Table.~\ref{tbl:W}.}
\label{fig:markov_system}
\end{figure}

\begin{table}[h!]
\begin{ruledtabular}
\caption{\label{tbl:W}
Non-zero elements of the transfer probability matrix $\hat W$}
\begin{tabular}{cccc}
${W}_{ij}$& Value & ${W}_{ij}$& Value\\
\hline
 ${W}_{11}$   & $1-\theta_1-\theta_3$   &${W}_{44}$ & $1-\theta_1-\theta_3$\\
 ${W}_{12}$&   $\theta_2$  & ${W}_{45}$  & $\theta_2$ \\
 ${W}_{13}$ & $\theta_4$ &${W}_{46}$&  $\theta_4$\\
 ${W}_{21}$   & $\theta_1$ & ${W}_{53}$&  $k_1$\\
 ${W}_{22}$ &  $1-\theta_2-k_3$   & ${W}_{54}$& $\theta_1$\\
 ${W}_{26}$ & $k_4$ & ${W}_{55}$   & $1-\theta_2-k_2$\\
 ${W}_{31}$& $\theta_3$ & ${W}_{62}$ & $k_3$\\
 ${W}_{33}$& $1-\theta_4-k_1$ & ${W}_{64}$ & $\theta_3$\\
 ${W}_{35}$ & $k_2$  & ${W}_{66}$ & $1-\theta_4-k_4$ \\
\end{tabular}
\end{ruledtabular}
\end{table}%

We construct a discrete-time 6-state Markov model (see Fig.~\ref{fig:markov_system}) to describe a binary-state sensor coupled to a 3-state environment. This is an extremely simple sensor in a simple environment, yet we can demonstrate that the sensor can perform multiplexing. The environment is simplified into 3-states: $E_1$, $E_2$, and $E_3$. At each time step, the environment can stay in the previous state (solid curved lines) or make a transition between state $E_1$ and $E_2$ or between $E_1$ and $E_3$ (solid straight lines), regardless of the state of the sensor. The 4 transitions, i.e., $E_1 \rightarrow E_2$, $E_2 \rightarrow E_1$, $E_1 \rightarrow E_3$, and $E_3 \rightarrow E_1$, are impacted by the nature of the environment, and their Markov transition probabilities are dictated by the environmental condition $\vtheta$. Thus we represent these four transition probabilities by $\vtheta=(\theta_1,\theta_2,\theta_3,\theta_4)$. The composite dynamics of the sensor and the environment can be captured by a 6-state discrete-time Markov Process \footnote{Notice that although the dynamics of the 6-state composite system is Markovian, its projection to the sensor's state space is no longer Markovian and have a history dependence.}: the sensor can also transition between states $s=0$ and $s=1$ while such transitions are coupled to the change of the environment. E.g., when the environment is in state $E_2$ and the sensor is at state $s=0$, the sensor can switch to state $s=1$, and causing the environment to change into $E_3$ due to the sensor--environment interactions (2 dashed straight lines). There are 4 such transitions (two from each dashed straight line), and their transition probabilities are $k_1$, $k_2$, $k_3$, $k_4$. 
As a consequence, $\vtheta=(\theta_1,\theta_2, \theta_3, \theta_4)$ characterizes the environment's kinetics and $\bm{k}=(k_1,k_2,k_3, k_4)$ characterizes the sensor's kinetics.  In summary, the composite system of the environment and the sensor with 6 possible composite states 1~--~``$(E_1,s_0)$", 2~--~``$(E_2,s_0)$", 3~--~``$(E_3,s_0)$", 4~--~``$(E_1,s_1)$", 5~--~``$(E_2,s_1)$", 6~--~``$(E_3,s_1)$" that evolves according to the Markov transition probability matrix $\hat W$ whose non-zero elements are listed in Table~\ref{tbl:W}. Notice that if the system is conditioned at state $j$, after a unit time, the probability of finding the system at state $i$ is the transfer probability, $W_{ij}$. Also, each element of the matrix (transition probability) must take the value between $0$ and $1$, and each column of the matrix must sum to 1.

\begin{figure*}
\includegraphics{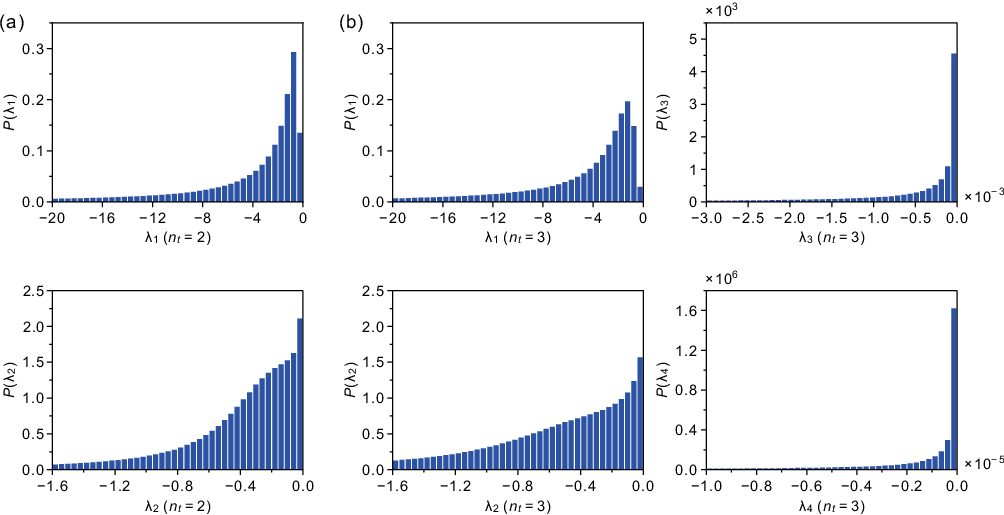}
\caption{The probability density of (a) the two non-zero eigenvalues ($\lambda_1 \leq \lambda_2<0$) for the llf's Hessian matrix obtained from $n_t=2$-point trajectories ($\lambda_3=\lambda_4=0$) and (b) the four non-zero eigenvalues for $n_t=3$-point trajectories ($\lambda_1\leq \lambda_2 \leq \lambda_3 \leq \lambda_4 <0$). Notice that there are 4 eigenvalues for both (a) and (b), and two eigenvalues for (a) are strictly zero. By comparing the 2 leading eigenvalues $\lambda_1$ and $\lambda_2$ for both cases $n_t=2$ and $n_t=3$, one finds that their distributions are similar. Thus, in this case, increasing the number of time steps in the trajectory can increase the number of independent variables that a sensor can sense but can not significantly improve the accuracy of sensing for variables that can already be sensed. }
\label{fig:histograms}
\end{figure*}
We present a rank-deficient Hessian analysis to characterize the number of independent environment variables that a receptor can sense. For a given sensor and given trajectory length, this number is evaluated by the rank of the corresponding Hessian matrix, $r_s$. Moreover, we have argued that the value of $r_s$ is bounded tightly by $r_s \leq d_{ir}(n_t,n_s)$, in Eq.~\ref{eq:tight}.

{\bf Multiplexing upper bound saturation.}
We conjecture that an arbitrarily designed sensor with the given network connectivity (Fig.~\ref{fig:markov_system}) can typically saturate the tightened bound, Eq.~\ref{eq:tight}. In other words, with a probability of approximately 1, a sensor with randomly chosen kinetic parameters corresponds to a Hessian matrix whose rank fully saturates the irreducible dimension of the trajectory probability space, $r_s=d_{ir}$.

For the simple Markov model, the sensor's trajectory probabilities can be used to obtain the Hessian matrices as described below. 
By randomly choosing a designated environmental condition $\vtheta$ and randomly choosing the design of the sensor $\bm{k}$, we solve for steady-state trajectory probabilities, $P_\vtheta[00]$, $P_\vtheta[01]$, $P_\vtheta[10]$ and $P_\vtheta[11]$ (see Appendix~\ref{Sec:AG}). Notice that we are interested in inferring the values of environmental parameters $\vtheta =(\theta_1,\theta_2,\theta_3,\theta_4)$, then we obtain a $4\times 4$ Hessian matrix for any randomly chosen composite system of sensor and environment. Then steady-state Hessian matrix of the llf for any $\vtheta$ and any $\bm{k}$ can be solved, and the $\{i,j\}-$th element of the Hessian matrix is expressed by Eq.~\ref{hessian_element} in Appendix~\ref{Sec:AG}.

Given the Hessian matrix of each composite system, one can perform the rd-MLE analysis and study the sensing capability of any randomly generated sensor in the 3-state environment. Here we randomly generate $10^7$ composite systems by sampling entries $\vtheta$ and $\bf k$ from the range $(0.0,0.2)$ with uniform probability. With these randomly generated composite systems, we can inquire about the histograms of the eigenvalues corresponding to their hessian matrix. 
In Fig.~\ref{fig:histograms}(a) we demonstrate for $n_t=2$ and in Fig.~\ref{fig:histograms}(b) for $n_t=3$, the histograms of the non-zero eigenvalues of Hessian for randomly generated composite systems ($\vtheta$ and $\bm{k}$). For $n_t=2$ we verify that almost all (with probability 1) randomly generated environments and randomly generated sensors give us only 2 negative eigenvalues, saturating the tightened upper bound of multiplexing ($r_s=d_{ir}(n_t=2,n_s=2)= (n_s^{n_t}-1)-1=2$); and for $n_t=3$, typically, there are $4$ negative eigenvalues, also saturating tightened upper bound of multiplexing ($r_s=d_{ir}(n_t=3,n_s=2)=(n_s^{n_t}-1)-3=4$). The plots in Fig.~\ref{fig:histograms} show the histogram of the eigenvalues ordered from the most negative to the least negative.

\section{Conclusion}
In this work, we demonstrate that from the information science perspective, a simple ligand-receptor sensor has the capacity to sense more than the ligand concentration. Multiple dimensions of environmental information, including temperature, ligand concentration, and flow speed, are embedded in the temporal trajectory of the receptor's state. A simple contour-line crossing technique demonstrates that one can simultaneously infer flow speed $v_x$ and ligand concentration $c$ from the statistics of 2-point trajectories of the sensor's state. In other words, a simple binary-state sensor can perform multiplexing. 

A general theoretical framework is presented for understanding multiplexing by noisy sensors in general. This theory predicts a theoretical upper bound on the multiplexing capacity. Specifically, the number of independent environmental variables that a sensor can infer is limited by $r_s\leq d_x(n_t,n_s) =n_s^{n_t}-1$, where $n_s$ is the number of states of the sensor and $n_t$ denotes the time-length of sensor's trajectory that the downstream signaling network can memorize.  We have also shown that the universal upper bound can be tightened by the inherent symmetries of the trajectory probability, whose analytical form cannot be obtained in general.

This work also presents a general tool (rd-MLE) to analyze the multiplexing capacity of a sensor in general. The rd-MLE predicts not only the multiplexing capacity $r_d$, but also provides an objective estimation of the sensor's sensitivities to multiple environmental variables.  
Moreover, by applying the rd-MLE to $10^7$ randomly generated sensors (described by Markov models), we conjecture that a randomly generated sensor can typically saturate the tightened upper bound of multiplexing without the need to carefully tune its parameters.

Combining information science and stochastic thermodynamics, this work provides an all-inclusive picture to describe the sensor-environment interaction, and provides a comprehensive and unbiased perspective on the sensor's ability to sense multi-dimensional environmental information.

\section*{Acknowledgements}
The authors appreciate financial support by funds from the National Science Foundation Grant DMR-2145256.  We would like to thank the University of North Carolina at Chapel Hill and the Research Computing group for providing computational resources and support that have contributed to these research results. ZL appreciates the encouragement of Tom Witten to include an analytically solvable model, ZL appreciates David Wolpert for pointing out that such behavior is called multiplexing in information science. The authors appreciate discussions and helpful comments on the manuscript from Tom Witten, Gary Pielak, Chase Slowey, Supraja Chittari, Zhongmin Zhang, Max Berkowitz, and Aaron Dinner.

\appendix
\section{Simulation details}
\label{Sec:AA}
\subsection{Ligand dynamics}
The simulation tube used in this study has dimensions $100 \times 20 \times 20$ in the $x$, $y$, and $z$ directions. There are periodic boundary conditions in the $x$-direction while reflective boundary conditions along the $y$- and $z$-directions. A single receptor is fixed at the center of the tube. The system is specified by multi-dimensional environmental information $\vtheta = (\mu,T,v_x)$, where the dimensionless $\mu=\ln(c/c^\standardstate)$ is the natural logarithm of the ratio between concentration $c$ and a reference concentration $c^\standardstate$, $T$ is temperature, and $v_x$ is the speed of media flow along $x$-direction. The ligands are modeled by Langevin dynamics according to Eq.~\ref{eq:Langevin}:
\begin{equation}
    \dot{\vec{r_i}}= v_x {\bf n}_x+ \frac{1}{{m}\gamma}\vec{F_i}(\vec{r_i})+\bm{\eta}(t)
\end{equation}
where $\vec{r}_i$ is the position of the $i$-th ligand molecule, $m$ is the ligand's mass which is set to 1, $\gamma$ is the friction coefficient set to 10, $v_x {\bf n}_x$ accounts for the background flow velocity of the media along the $x$-axis (${\bf n}_x$), $\vec{F}(\vec{r})$ is the total deterministic force experienced by each ligand molecule (in a detailed simulation, this force accounts for the sum of pairwise interactions between ligand pairs and the ligand-receptor attraction), and $\bm{\eta}$ is the Gaussian noise characterized by $\langle\eta_i(t)\eta_j(t')\rangle = 2\gamma k_BT\delta_{ij}\delta(t-t')$, where $k_B$ is the Boltzmann constant. The unit is chosen such that $k_B=1$. The ligand--ligand interaction can either be ignored (ideal solution) or it can adopt a short-range repulsive WCA potential \cite{weeks1971perturbation}:
\begin{equation}
    U_{l,l}=\begin{cases}
			4\epsilon\bigg(\big(\frac{\sigma}{d_{l,l}}\big)^{12}-\big(\frac{\sigma}{d_{l,l}}\big)^6\bigg)+\epsilon, & \text{if $d_{l,l}\leq 2^{1/6}\sigma$}\\
            0, & \text{otherwise}
		 \end{cases}
	\label{eq:L-L interaction}
\end{equation}
Here $d_{l,l}$ is the distance between two ligand particles, and the cutoff distance is equal to $2^{1/6}\sigma$. The parameter values were fixed to $\epsilon = 0.01$ and $\sigma = 1.5$.  
In the simulation, we choose the integrator time step as $dt = 0.01$.
Taking the ideal solution limit by ignoring the ligand--ligand interaction can significantly reduce the computational time yet the result is qualitatively the same as that obtained with the WCA potential. See the comparison in Appendix~\ref{Sec:AB}.

\subsection{Receptor dynamics}
The interaction between the receptor and the ligand, and in particular the state-dynamics of the receptor is modeled by a stochastic transition process. At each simulation time step, the receptor can bind or unbind with a ligand, and its probability is determined by the environment and the ligand's state at the simulation snapshot. 
At any time step, if the receptor is empty, then there is a probability for binding, which is impacted by the number of free ligands around its vicinity. 
A ligand within a cutoff radius $r_c$ of $2.249$ from the receptor is considered as within the binding vicinity. The binding probability rate of the receptor is expressed as 
\begin{equation}
k_+= n_l e^{-\beta (E_b-E_{\rm off})}
\end{equation}
where $n_l$ is the number of ligands within the cutoff radius, $E_{\rm off}$ is the energy of an empty receptor, and $E_b$ is the energy barrier for the binding-unbinding transition. $n_l$ is a fluctuating dimensionless quantity related to a fluctuating local concentration $c'$ as
\begin{equation}
\label{eq:nl}
    n_l=\frac{4\pi r_c^3}{3 V_{\text{tube}}}\frac{c'}{c^\standardstate}
\end{equation}
where $V_{\text{tube}}$ is the volume of the whole simulation tube, $r_c$ is the cutoff radius around the receptor, and we have chosen a reference concentration (unit concentration) $c^\standardstate$ defined by the situation where there is only one ligand molecule within the whole tube. To estimate binding rate $k_+$ in the Langevin dynamics simulation, we do not explicitly evaluate $c'$ but only need to count $n_l$, which is the number of ligand particles within the cutoff radius. At each time step, if the receptor is empty, the probability of ``on'' is 
\begin{equation}
    P_{+}=R_{\rm on} \cdot dt
\end{equation}
where $dt=0.01$ is the simulation time step.
When the binding occurs, the number of free ligands in the local environment (and the whole simulation tube) reduces by 1.

If the receptor is already bound with one ligand, then the unbinding event could occur, resulting in one ligand returning back into the bath. Considering the effect of the media flow speed $v_x$ and the friction coefficient that each ligand experiences $\gamma$, the unbinding occurs at the rate of 
\begin{equation}
 k_{-}= e^{-\beta (E_b-E_{\rm on}-\gamma\alpha v_x)}
\end{equation}
where $E_{\rm on}$ is the energy of the receptor when it is bound with the ligand set to be equal to $-1.0$, and the term $\gamma\alpha v_x$ is the frictional work done by the flowing media to assist the ligand's dissociation from the receptor. Here we have assumed that the ligand needs to displace from the receptor by a distance $\alpha = 0.1$ to be considered as unbound from the ligand. However, the dissipative term can be ignored and will not qualitatively change the sensor's ability to sense multiple environmental variables (as illustrated in Fig. \ref{fig:nodrag}). 
At each time step, if the receptor is bound with a ligand, the probability of ``off'' is 
\begin{equation}
    P_{-}=R_{\rm off} \cdot dt
\end{equation}
where $dt=0.01$ is the simulation time step. When unbinding occurs, one ligand is returned to be the free ligand in the solution.

\section{Two approaches to Ligand--ligand interactions}
\label{Sec:AB}
As mentioned in Appendix A, the results shown in Figs.~\ref{fig:cross}, \ref{fig:mu_beta} and \ref{fig:beta_vx} are obtained through simulations where the ligands were considered as free particles and did not interact with each other. 

\begin{figure*}
\centering
\includegraphics[width=\textwidth]{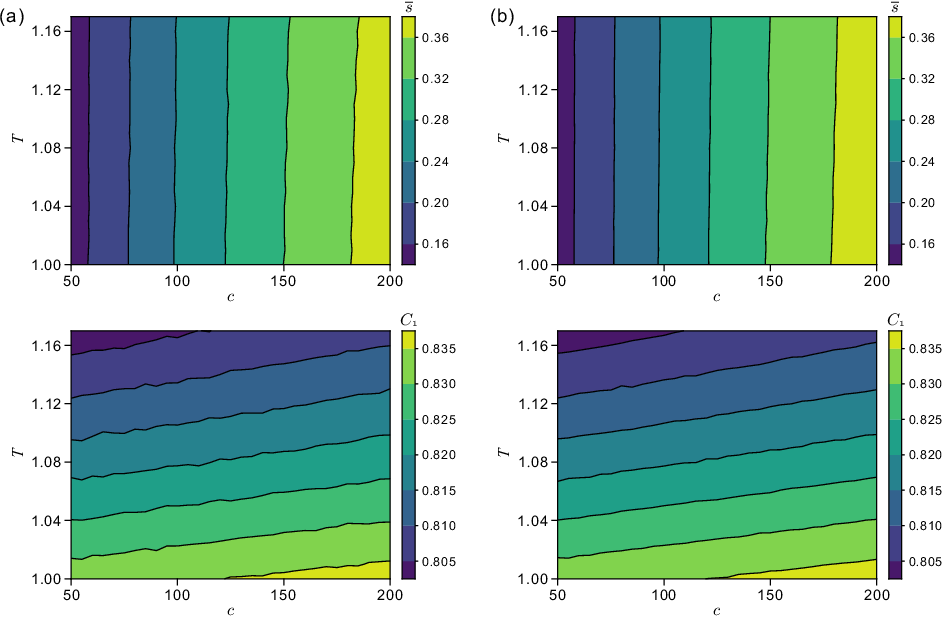}
\caption{The contour plots of $\bar{s}$ and $C_1$ plotted against the variables $c$ and $T$ when the ligand particles were modeled using (a) the WCA potential and (b) free particles. The simulations were performed for 100 million time-steps with 20 repetitions and 10 billion time-steps in the case of (a) and (b), respectively.}
\label{fig:WCA_T_c}
\end{figure*}

\begin{figure*}
\includegraphics[width=\textwidth]{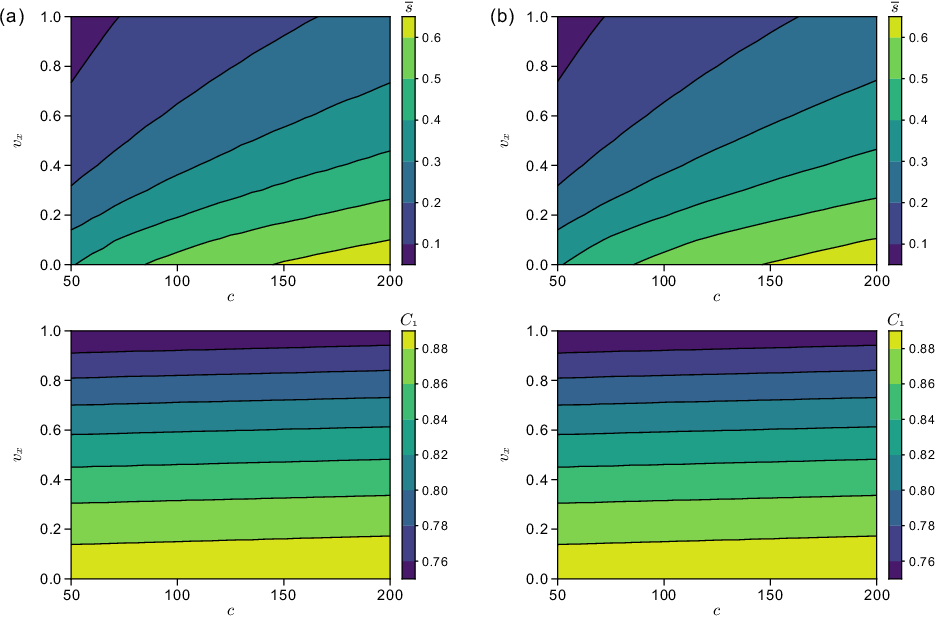}
\caption{The contour plots of $\bar{s}$ and $C_1$ plotted against the variables $c$ and $v_x$ when the ligand particles were modeled using (a) the WCA potential and (b) free particles. The simulations were performed for 100 million time-steps with 20 repetitions and 10 billion time-steps in the case of (a) and (b), respectively.}
\label{fig:WCA_v_c}
\end{figure*}

\begin{figure*}
\includegraphics[width=\textwidth]{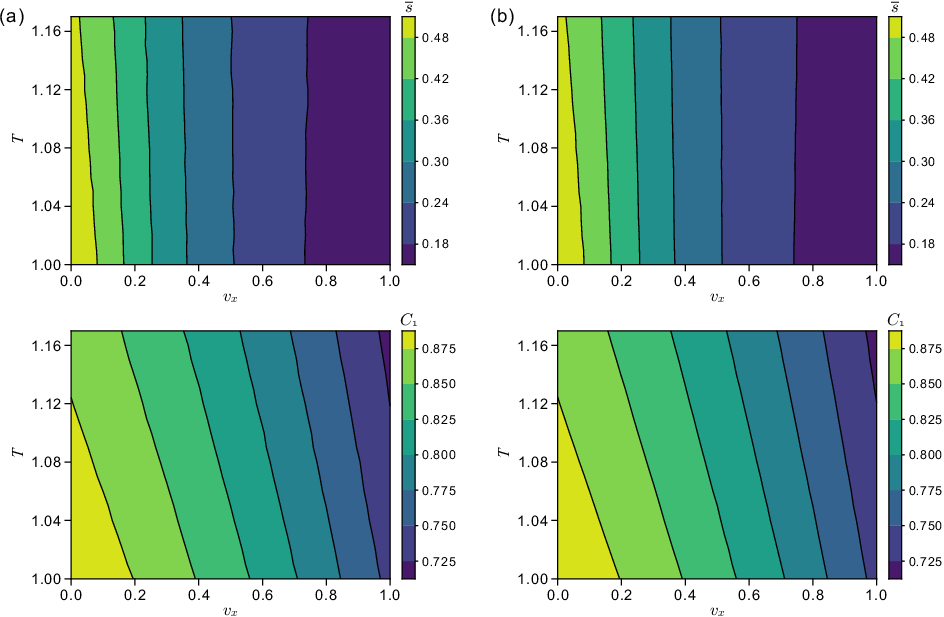}
\caption{The contour plots of $\bar{s}$ and $C_1$ plotted against the variables $v_x$ and $T$ when the ligand particles were modeled using (a) the WCA potential and (b) free particles. The simulations were performed for 100 million time-steps with 20 repetitions and 10 billion time-steps in the case of (a) and (b), respectively.}
\label{fig:WCA_T_vx}
\end{figure*}

To make sure the result is independent of the choice of model for the ligand--ligand interactions, we also performed simulations where the ligands interacted with each other through the WCA potential given by Eq.~\ref{eq:L-L interaction}. These results are summarized and compared with similar simulations performed for ideal ligand solutions (no ligand--ligand interaction) in Figs.~\ref{fig:WCA_T_c}, \ref{fig:WCA_v_c}, and \ref{fig:WCA_T_vx}. These contour plots clearly show that the results remain qualitatively unchanged regardless of whether the interactions between the ligand particles are considered WCA repulsions or ignored. It demonstrates that our result is robust against details of the specific model. 

\section{Multiplexing in the absence of frictional drag}
\label{Sec:AC}
\begin{figure}
\includegraphics{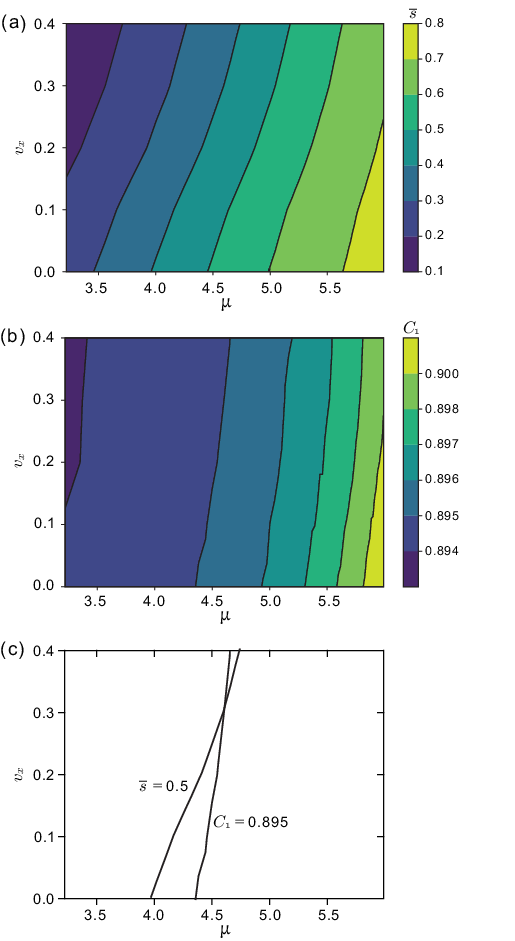}
\caption{The contour plots of $\bar{s}$ and $C_1$ plotted against the variables $\mu$ and $v_x$ when the frictional drag term is eliminated from the binding kinetics of the receptor.}
\label{fig:nodrag}
\end{figure}

As described in Appendix A, the unbinding kinetics of the sensor are impacted by a frictional drag that is caused due to the flow in the system. However, to verify that the system can perform multiplexing even without the explicit effect of the flow on the receptor's dynamics, we performed simulations without this term in the unbinding kinetics. As shown in Fig.~\ref{fig:nodrag}, the receptor can still sense flow speed and concentration simultaneously and hence perform multiplexing. This implies that multiplexing demonstrated in this work is independent of the explicit dynamics of the receptor.

\section{Contour plots}
\label{Sec:AD}
\begin{figure}
\includegraphics{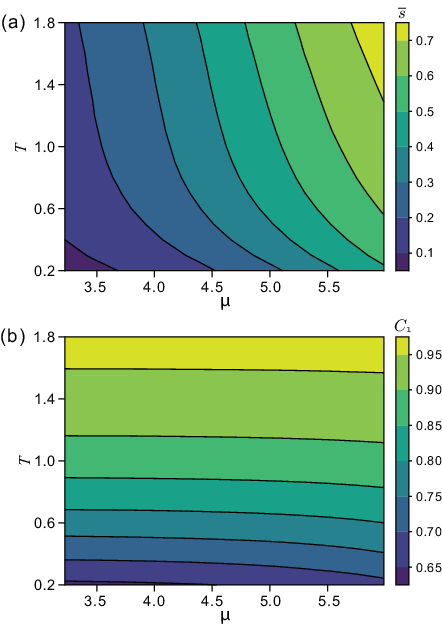}
\caption{The contour plots of $\bar{s}$ and $C_1$ plotted against the variables $\mu$ and temperature $T$.}
\label{fig:mu_beta}
\end{figure}

\begin{figure}
\includegraphics{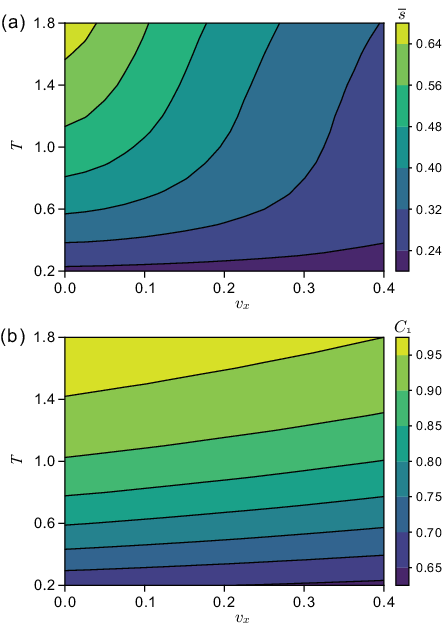}
\caption{The contour plots of $\bar{s}$ and $C_1$ plotted against the variables $v_x$ and temperature $T$.}
\label{fig:beta_vx}
\end{figure}

We presented the contour plots for the case where $T = 1.0$ was fixed and $\mu \in [\ln 25, \ln 398]$ $v_x\in[0.0,0.4]$ in Fig.~\ref{fig:cross}. There we have shown a contour-line crossing technique: if the values of $\bar{s}$ and $C_1$ are known, then one can simultaneously resolve the values for the two variables $\mu$ and $v_x$. Here, we present the contour plots of the other two combinations of the variables: $\mu \in [\ln 25, \ln 398]$ and $T\in[0.2,1.8]$, with a fixed $v_x=0.2$, shown in Fig.~\ref{fig:mu_beta}; and $T\in[0.2,1.8]$ and $v_x\in[0.0,0.4]$ with a fixed $\mu = \ln 100$, as shown in Fig.~\ref{fig:beta_vx}. It is clear from these contour plots that the contour lines corresponding to $\bar{s}$ and $C_1$ respectively are not parallel to each other, except for a small region in Fig.~\ref{fig:beta_vx}. Thus for most range of parameters, the contour-line crossing technique can be used to simultaneously infer two independent environmental values from two-point trajectory statistics. 


\section{Symmetries}
\label{Sec:AE}
\subsection{2-point trajectories}
When $n_t=2$, the possible trajectories are $S\in\{00,01,10,11\}$. For a sensor that is sensing for a long enough time, the probability $P_\vtheta[01]=P_\vtheta[10]$. This is true for a binary-state sensor: over a long time, regardless of the length of the trajectory, the total binding events (01) must be almost equal to the total unbinding events (10) and their difference's absolute value must be smaller than or equal to 1. Averaged over a long trajectory, we must have 
\begin{equation}
    P_\vtheta[01]=P_\vtheta[10]
\end{equation}

\subsection{3-point trajectories}
When $n_t=3$, the possible trajectories are $S\in\{000,001,010,011,100,101,110,111\}$.
In this case, the symmetries for a binary state sensor, similar to the argument from above can be shown to be 3-fold:
\begin{align}
    P_\vtheta[001]&=P_\vtheta[100]\\
    P_\vtheta[011]&=P_\vtheta[110]\\ P_\vtheta[001]+P_\vtheta[101]&=P_\vtheta[010]+P_\vtheta[110]
\end{align}
The symmetries listed above will be discussed in detail in future work focusing on the optimal design of good sensors for multiplexing.

\section{Numerical simulations for obtaining the Hessian matrix}
\label{Sec:AF}
The Hessian of the ligand-receptor demonstrated in Eq.~\ref{eq:hessian} was obtained through the simulation with the following parameters: $n_s = 2$ (binary-state sensor), the length of a trajectory $N=10^{10}$, time step $dt=0.01$, and  $(\mu^*,T^*,v_x^*)= (\ln{100}, 1.0, 0.2)$. The Hessian involves second-order derivatives, which are computed as finite differences with $\delta\theta = (4.8459\%$ of $\mu^*, 5\%$ of $T^*, 5\%$ of $v_x^*)$. The result was acquired for $n_t = 2$ with only 2-point trajectories, and the trajectories are taken as $S=(s(t),~s(t+100 dt))$.

\section{Analysis of Markov model}
\label{Sec:AG}
\subsection{Solution to the Trajectory Probabilities}

For the Markov model described in Sec. IV.B, the discrete-time evolution of the probability of the composite environment-sensor system $\vec p$ can be written as
\begin{equation}
    \vec p(t_{i+1}) = \hat W~\vec p(t_i)
\end{equation}
where $\vec p(t=t_i)$ is a $6$-by-$1$ probability vector for states 1 to 6 at time $t=t_i$ and $\hat W$ is the transfer probability matrix described by the transition rates listed in Table.~\ref{tbl:W}. At any time, we can write the probability of the sensor as $P[0]= p_1+p_2+p_3$ and $P[1]= p_4+p_5+p_6$. Below, we obtain the probability of trajectories rather than probabilities of states.

At stationary environmental condition $\vtheta$ and a given design of the sensor $\bf k$, $\hat{W}$ is time-homogeneous and the system could reach non-equilibrium steady state (NESS) with probability over the 6-state state space denoted by $\vec {p}_{ss}$: 
\begin{equation}
    \hat W~ \vec{p}_{ss}=\vec{p}_{ss}
\end{equation}
Obtaining $\vec p_{ss}$ allows us to obtain the discrete-time trajectory probabilities of the sensor's state. Notice that among the 6 states, the first 3 states correspond to the sensor in the state ``u" ($s=0$) and the last 3 states correspond to ``b" ($s=1$). For convenience let us define the following vectors and matrices to assist the coarse-graining from the 6-state probability vector to probabilities of the sensor's binary state:
\begin{align}
    \vec u &= (1,1,1,0,0,0) \\
    \vec b &= (0,0,0,1,1,1) \\
    \hat D_u &= \begin{pmatrix} 1 & 0 & 0 & 0 & 0 & 0 \\ 0 & 1 & 0 & 0 & 0 & 0 \\ 0 & 0 & 1 & 0 & 0 & 0 \\ 0 & 0 & 0 & 0 & 0 & 0 \\ 0 & 0 & 0 & 0 & 0 & 0 \\ 0 & 0 & 0 & 0 & 0 & 0 \\ \end{pmatrix}\\
    \hat D_b &= \begin{pmatrix} 0 & 0 & 0 & 0 & 0 & 0 \\ 0 & 0 & 0 & 0 & 0 & 0 \\ 0 & 0 & 0 & 0 & 0 & 0 \\ 0 & 0 & 0 & 1 & 0 & 0 \\ 0 & 0 & 0 & 0 & 1 & 0 \\ 0 & 0 & 0 & 0 & 0 & 1 \\ \end{pmatrix}
\end{align}

First, let us focus on the trajectory probabilities of the sensor's 2-point trajectories ($n_t=2$). Here the sensor's state at time $t$ and at time $t+1$ gives 4 possible trajectories $S \in\{00,01,10,11\}$. At NESS, with a given environmental condition $\vtheta$, the trajectory probability can be written as
\begin{align}
    P_\vtheta[00] &= \vec u \cdot (\hat{W}~(\hat D_u ~ \vec{p}_{ss})) \\
    P_\vtheta[01] &= \vec b \cdot (\hat{W}~(\hat D_u ~ \vec{p}_{ss})) \\
    P_\vtheta[10] &= \vec u \cdot (\hat{W}~(\hat D_b ~ \vec{p}_{ss})) \\
    P_\vtheta[11] &= \vec b \cdot (\hat{W}~(\hat D_b ~ \vec{p}_{ss})) 
\end{align}
Then for $n_t=3$, where the sensor's state at times $t$, $t+1$, and $t+2$ forms 3-point trajectories: $S \in \{000,001,010,011, 100,101,110,111\} $. One can write the trajectory probabilities as:
\begin{align}
    P_\vtheta[000] &= \vec u \cdot (\hat{W}~(\hat D_u~(\hat{W}~(\hat D_u ~ \vec{p}_{ss})))) \\
    P_\vtheta[001] &= \vec u \cdot(\hat{W}~(\hat D_u~(\hat{W}~(\hat D_b ~ \vec{p}_{ss})))) \\
    P_\vtheta[010] &= \vec u \cdot(\hat{W}~(\hat D_b~(\hat{W}~(\hat D_u ~ \vec{p}_{ss})))) \\
    P_\vtheta[011] &= \vec u \cdot(\hat{W}~(\hat D_b~(\hat{W}~(\hat D_b ~ \vec{p}_{ss})))) \\
    P_\vtheta[100] &= \vec b \cdot(\hat{W}~(\hat D_u~(\hat{W}~(\hat D_u ~ \vec{p}_{ss})))) \\
    P_\vtheta[101] &= \vec b \cdot(\hat{W}~(\hat D_u~(\hat{W}~(\hat D_b ~ \vec{p}_{ss})))) \\
    P_\vtheta[110] &= \vec b \cdot(\hat{W}~(\hat D_b~(\hat{W}~(\hat D_u ~ \vec{p}_{ss})))) \\
    P_\vtheta[111] &= \vec b \cdot(\hat{W}~(\hat D_b~(\hat{W}~(\hat D_b ~ \vec{p}_{ss}))))
\end{align}
The analytical expression is further used to analytically obtain the log-likelihood function (llf) and Hessian matrices by methods described below. These expressions are too long to be displayed in terms of matrix elements of $\hat W$, and were evaluated analytically using Mathematica.

\subsection{Rank-deficient MLE analysis}

Consider a fixed environmental variable $\vtheta^*$, we assume that accumulate $m$ independent trajectories of the sensor's state, $\{S^{*1},S^{*2},\cdots S^{*m}\}$. One can perform a maximum likelihood estimation (MLE) of $\vtheta^*$ where the log-likelihood function (llf) of environmental variables $\vtheta$ under sufficient sampling $m\gg1$ can be written as
\begin{equation}
\label{eq:llf}
l(\vtheta) = \sum_{S} P_{\vthetas}[S]\ln P_{\vtheta}[S]
\end{equation}
where $P_{\vthetas}[S]$'s are probability to observe a trajectory $S$ at the environment condition $\vthetas$, and $P_{\vtheta}[S]$'s are the trajectory probabilities for any arbitrary parameter $\vtheta$. The maximum likelihood estimator's job is to estimate the true condition of the environment ($\vthetas$) by varying the value of $\vtheta$ and maximizing the llf $l(\vtheta)$. Then $\vtheta_{est}=argmax \{l (\vtheta)\}$ is the best estimation of the environmental condition $\vthetas$.  In the ideal case of full multiplexing, where all environmental variables can be simultaneously sensed from the trajectories, $\vtheta_{est} \rightarrow \vthetas$ in the limit of sufficient sampling. 

The multiplexing ability is then determined by the geometry of $l(\vtheta)$ around $\vthetas$. If $l(\vtheta)$ is strictly concave, then full multiplexing can be achieved. Here the capability of multiplexing around $\vthetas$ is thus determined by the rank of the Hessian matrix of $l(\vtheta)$ evaluated at $\vthetas$. Below let us first analytically obtain the Hessian matrix.
The $ij$-th element of the Hessian is expressed as: 
\begin{equation}
    H_{ij}=\frac{\partial^2 l(\vtheta)}{\partial \theta_i \partial \theta_j}
\end{equation}
Given the expression for the llf under abundant sampling, Eq.~\ref{eq:llf}, the first derivative of the $l(\vtheta)$ can be expressed as
\begin{equation}
    \frac{\partial l(\vtheta)}{\partial \theta_j} = \sum_{S} P_{\vtheta^*}[S]\frac{1}{P_{\vtheta}[S]}\frac{\partial P_{\vtheta}[S]}{\partial\theta_j}\bigg |_{\vtheta^*}
\end{equation} 
Then the $ij$-th element of the Hessian is
\begin{equation}
    \frac{\partial ^2 l(\vtheta)}{\partial \theta_i \partial\theta_j} = \sum_{S} \bigg\{ \frac{\partial ^2 P_{\vtheta}[S]}{\partial \theta_i \partial \theta_j} -\frac{1}{P_{\vtheta}[S]} \frac{\partial P_{\vtheta}[S]}{\partial \theta_i} \frac{\partial P_{\vtheta}[S]}{\partial \theta_j}\bigg\}\bigg |_{\vtheta^*}
    \label{eq:second_derivative}
\end{equation}
The first term in Eq.~\ref{eq:second_derivative} vanishes due to normalization, and hence we obtain
\begin{equation}
\label{hessian_element}
H_{ij}=-\sum_{S}\frac{1}{P_\vtheta [S]}\frac{\partial P_\vtheta [S]}{\partial\theta_i}\frac{\partial P_\vtheta [S]}{\partial\theta_j}\bigg |_{\vtheta^*}
\end{equation}
In the specific example of the 6-state-Markov model, we have declared that there are $4$ independent environmental variables $\theta_1$, $\theta_2$, $\theta_3$, and $\theta_4$, and correspondingly the Hessian is a $4\times4$ matrices for both cases of $n_t=2$ and $n_t=3$. The Hessian is rank-deficient when its rank is lower than $4$.

\section{Codes}
\label{Sec:AH}
The numerical code written in Julia language will be made publicly available at GitHub: \href{https://github.com/ljpotential/StochasticSensoryReceptor.jl}{Stochastic Sensory Receptor}

\bibliography{apssamp}

\begin{thebibliography}{35}%
\makeatletter
\providecommand \@ifxundefined [1]{%
 \@ifx{#1\undefined}
}%
\providecommand \@ifnum [1]{%
 \ifnum #1\expandafter \@firstoftwo
 \else \expandafter \@secondoftwo
 \fi
}%
\providecommand \@ifx [1]{%
 \ifx #1\expandafter \@firstoftwo
 \else \expandafter \@secondoftwo
 \fi
}%
\providecommand \natexlab [1]{#1}%
\providecommand \enquote  [1]{``#1''}%
\providecommand \bibnamefont  [1]{#1}%
\providecommand \bibfnamefont [1]{#1}%
\providecommand \citenamefont [1]{#1}%
\providecommand \href@noop [0]{\@secondoftwo}%
\providecommand \href [0]{\begingroup \@sanitize@url \@href}%
\providecommand \@href[1]{\@@startlink{#1}\@@href}%
\providecommand \@@href[1]{\endgroup#1\@@endlink}%
\providecommand \@sanitize@url [0]{\catcode `\\12\catcode `\$12\catcode
  `\&12\catcode `\#12\catcode `\^12\catcode `\_12\catcode `\%12\relax}%
\providecommand \@@startlink[1]{}%
\providecommand \@@endlink[0]{}%
\providecommand \url  [0]{\begingroup\@sanitize@url \@url }%
\providecommand \@url [1]{\endgroup\@href {#1}{\urlprefix }}%
\providecommand \urlprefix  [0]{URL }%
\providecommand \Eprint [0]{\href }%
\providecommand \doibase [0]{https://doi.org/}%
\providecommand \selectlanguage [0]{\@gobble}%
\providecommand \bibinfo  [0]{\@secondoftwo}%
\providecommand \bibfield  [0]{\@secondoftwo}%
\providecommand \translation [1]{[#1]}%
\providecommand \BibitemOpen [0]{}%
\providecommand \bibitemStop [0]{}%
\providecommand \bibitemNoStop [0]{.\EOS\space}%
\providecommand \EOS [0]{\spacefactor3000\relax}%
\providecommand \BibitemShut  [1]{\csname bibitem#1\endcsname}%
\let\auto@bib@innerbib\@empty
\bibitem [{\citenamefont {Antebi}\ \emph
  {et~al.}(2017{\natexlab{a}})\citenamefont {Antebi}, \citenamefont {Linton},
  \citenamefont {Klumpe}, \citenamefont {Bintu}, \citenamefont {Gong},
  \citenamefont {Su}, \citenamefont {McCardell},\ and\ \citenamefont
  {Elowitz}}]{antebi2017combinatorial}%
  \BibitemOpen
  \bibfield  {author} {\bibinfo {author} {\bibfnamefont {Y.~E.}\ \bibnamefont
  {Antebi}}, \bibinfo {author} {\bibfnamefont {J.~M.}\ \bibnamefont {Linton}},
  \bibinfo {author} {\bibfnamefont {H.}~\bibnamefont {Klumpe}}, \bibinfo
  {author} {\bibfnamefont {B.}~\bibnamefont {Bintu}}, \bibinfo {author}
  {\bibfnamefont {M.}~\bibnamefont {Gong}}, \bibinfo {author} {\bibfnamefont
  {C.}~\bibnamefont {Su}}, \bibinfo {author} {\bibfnamefont {R.}~\bibnamefont
  {McCardell}},\ and\ \bibinfo {author} {\bibfnamefont {M.~B.}\ \bibnamefont
  {Elowitz}},\ }\bibfield  {title} {\bibinfo {title} {Combinatorial signal
  perception in the bmp pathway},\ }\href@noop {} {\bibfield  {journal}
  {\bibinfo  {journal} {Cell}\ }\textbf {\bibinfo {volume} {170}},\ \bibinfo
  {pages} {1184} (\bibinfo {year} {2017}{\natexlab{a}})}\BibitemShut {NoStop}%
\bibitem [{\citenamefont {Antebi}\ \emph
  {et~al.}(2017{\natexlab{b}})\citenamefont {Antebi}, \citenamefont
  {Nandagopal},\ and\ \citenamefont {Elowitz}}]{antebi2017operational}%
  \BibitemOpen
  \bibfield  {author} {\bibinfo {author} {\bibfnamefont {Y.~E.}\ \bibnamefont
  {Antebi}}, \bibinfo {author} {\bibfnamefont {N.}~\bibnamefont {Nandagopal}},\
  and\ \bibinfo {author} {\bibfnamefont {M.~B.}\ \bibnamefont {Elowitz}},\
  }\bibfield  {title} {\bibinfo {title} {An operational view of intercellular
  signaling pathways},\ }\href@noop {} {\bibfield  {journal} {\bibinfo
  {journal} {Current opinion in systems biology}\ }\textbf {\bibinfo {volume}
  {1}},\ \bibinfo {pages} {16} (\bibinfo {year}
  {2017}{\natexlab{b}})}\BibitemShut {NoStop}%
\bibitem [{\citenamefont {Lestas}\ \emph {et~al.}(2010)\citenamefont {Lestas},
  \citenamefont {Vinnicombe},\ and\ \citenamefont
  {Paulsson}}]{lestas2010fundamental}%
  \BibitemOpen
  \bibfield  {author} {\bibinfo {author} {\bibfnamefont {I.}~\bibnamefont
  {Lestas}}, \bibinfo {author} {\bibfnamefont {G.}~\bibnamefont {Vinnicombe}},\
  and\ \bibinfo {author} {\bibfnamefont {J.}~\bibnamefont {Paulsson}},\
  }\bibfield  {title} {\bibinfo {title} {Fundamental limits on the suppression
  of molecular fluctuations},\ }\href@noop {} {\bibfield  {journal} {\bibinfo
  {journal} {Nature}\ }\textbf {\bibinfo {volume} {467}},\ \bibinfo {pages}
  {174} (\bibinfo {year} {2010})}\BibitemShut {NoStop}%
\bibitem [{\citenamefont {Hinczewski}\ and\ \citenamefont
  {Thirumalai}(2014)}]{hinczewski2014cellular}%
  \BibitemOpen
  \bibfield  {author} {\bibinfo {author} {\bibfnamefont {M.}~\bibnamefont
  {Hinczewski}}\ and\ \bibinfo {author} {\bibfnamefont {D.}~\bibnamefont
  {Thirumalai}},\ }\bibfield  {title} {\bibinfo {title} {Cellular signaling
  networks function as generalized wiener-kolmogorov filters to suppress
  noise},\ }\href@noop {} {\bibfield  {journal} {\bibinfo  {journal} {Physical
  Review X}\ }\textbf {\bibinfo {volume} {4}},\ \bibinfo {pages} {041017}
  (\bibinfo {year} {2014})}\BibitemShut {NoStop}%
\bibitem [{\citenamefont {Berg}\ and\ \citenamefont
  {Purcell}(1977)}]{berg1977physics}%
  \BibitemOpen
  \bibfield  {author} {\bibinfo {author} {\bibfnamefont {H.~C.}\ \bibnamefont
  {Berg}}\ and\ \bibinfo {author} {\bibfnamefont {E.~M.}\ \bibnamefont
  {Purcell}},\ }\bibfield  {title} {\bibinfo {title} {Physics of
  chemoreception},\ }\href@noop {} {\bibfield  {journal} {\bibinfo  {journal}
  {Biophysical journal}\ }\textbf {\bibinfo {volume} {20}},\ \bibinfo {pages}
  {193} (\bibinfo {year} {1977})}\BibitemShut {NoStop}%
\bibitem [{\citenamefont {Bialek}\ and\ \citenamefont
  {Setayeshgar}(2005)}]{bialek2005physical}%
  \BibitemOpen
  \bibfield  {author} {\bibinfo {author} {\bibfnamefont {W.}~\bibnamefont
  {Bialek}}\ and\ \bibinfo {author} {\bibfnamefont {S.}~\bibnamefont
  {Setayeshgar}},\ }\bibfield  {title} {\bibinfo {title} {Physical limits to
  biochemical signaling},\ }\href@noop {} {\bibfield  {journal} {\bibinfo
  {journal} {Proceedings of the National Academy of Sciences}\ }\textbf
  {\bibinfo {volume} {102}},\ \bibinfo {pages} {10040} (\bibinfo {year}
  {2005})}\BibitemShut {NoStop}%
\bibitem [{\citenamefont {Kaizu}\ \emph {et~al.}(2014)\citenamefont {Kaizu},
  \citenamefont {De~Ronde}, \citenamefont {Paijmans}, \citenamefont
  {Takahashi}, \citenamefont {Tostevin},\ and\ \citenamefont
  {Ten~Wolde}}]{kaizu2014berg}%
  \BibitemOpen
  \bibfield  {author} {\bibinfo {author} {\bibfnamefont {K.}~\bibnamefont
  {Kaizu}}, \bibinfo {author} {\bibfnamefont {W.}~\bibnamefont {De~Ronde}},
  \bibinfo {author} {\bibfnamefont {J.}~\bibnamefont {Paijmans}}, \bibinfo
  {author} {\bibfnamefont {K.}~\bibnamefont {Takahashi}}, \bibinfo {author}
  {\bibfnamefont {F.}~\bibnamefont {Tostevin}},\ and\ \bibinfo {author}
  {\bibfnamefont {P.~R.}\ \bibnamefont {Ten~Wolde}},\ }\bibfield  {title}
  {\bibinfo {title} {The berg-purcell limit revisited},\ }\href@noop {}
  {\bibfield  {journal} {\bibinfo  {journal} {Biophysical journal}\ }\textbf
  {\bibinfo {volume} {106}},\ \bibinfo {pages} {976} (\bibinfo {year}
  {2014})}\BibitemShut {NoStop}%
\bibitem [{\citenamefont {Mora}\ and\ \citenamefont
  {Nemenman}(2019)}]{mora2019physical}%
  \BibitemOpen
  \bibfield  {author} {\bibinfo {author} {\bibfnamefont {T.}~\bibnamefont
  {Mora}}\ and\ \bibinfo {author} {\bibfnamefont {I.}~\bibnamefont
  {Nemenman}},\ }\bibfield  {title} {\bibinfo {title} {Physical limit to
  concentration sensing in a changing environment},\ }\href@noop {} {\bibfield
  {journal} {\bibinfo  {journal} {Physical review letters}\ }\textbf {\bibinfo
  {volume} {123}},\ \bibinfo {pages} {198101} (\bibinfo {year}
  {2019})}\BibitemShut {NoStop}%
\bibitem [{\citenamefont {Carballo-Pacheco}\ \emph {et~al.}(2019)\citenamefont
  {Carballo-Pacheco}, \citenamefont {Desponds}, \citenamefont {Gavrilchenko},
  \citenamefont {Mayer}, \citenamefont {Prizak}, \citenamefont {Reddy},
  \citenamefont {Nemenman},\ and\ \citenamefont {Mora}}]{carballo2019receptor}%
  \BibitemOpen
  \bibfield  {author} {\bibinfo {author} {\bibfnamefont {M.}~\bibnamefont
  {Carballo-Pacheco}}, \bibinfo {author} {\bibfnamefont {J.}~\bibnamefont
  {Desponds}}, \bibinfo {author} {\bibfnamefont {T.}~\bibnamefont
  {Gavrilchenko}}, \bibinfo {author} {\bibfnamefont {A.}~\bibnamefont {Mayer}},
  \bibinfo {author} {\bibfnamefont {R.}~\bibnamefont {Prizak}}, \bibinfo
  {author} {\bibfnamefont {G.}~\bibnamefont {Reddy}}, \bibinfo {author}
  {\bibfnamefont {I.}~\bibnamefont {Nemenman}},\ and\ \bibinfo {author}
  {\bibfnamefont {T.}~\bibnamefont {Mora}},\ }\bibfield  {title} {\bibinfo
  {title} {Receptor crosstalk improves concentration sensing of multiple
  ligands},\ }\href@noop {} {\bibfield  {journal} {\bibinfo  {journal}
  {Physical Review E}\ }\textbf {\bibinfo {volume} {99}},\ \bibinfo {pages}
  {022423} (\bibinfo {year} {2019})}\BibitemShut {NoStop}%
\bibitem [{\citenamefont {Hu}\ \emph {et~al.}(2010)\citenamefont {Hu},
  \citenamefont {Chen}, \citenamefont {Rappel},\ and\ \citenamefont
  {Levine}}]{hu2010physical}%
  \BibitemOpen
  \bibfield  {author} {\bibinfo {author} {\bibfnamefont {B.}~\bibnamefont
  {Hu}}, \bibinfo {author} {\bibfnamefont {W.}~\bibnamefont {Chen}}, \bibinfo
  {author} {\bibfnamefont {W.-J.}\ \bibnamefont {Rappel}},\ and\ \bibinfo
  {author} {\bibfnamefont {H.}~\bibnamefont {Levine}},\ }\bibfield  {title}
  {\bibinfo {title} {Physical limits on cellular sensing of spatial
  gradients},\ }\href@noop {} {\bibfield  {journal} {\bibinfo  {journal}
  {Physical review letters}\ }\textbf {\bibinfo {volume} {105}},\ \bibinfo
  {pages} {048104} (\bibinfo {year} {2010})}\BibitemShut {NoStop}%
\bibitem [{\citenamefont {Fran{\c{c}}ois}\ and\ \citenamefont
  {Zilman}(2019)}]{franccois2019physical}%
  \BibitemOpen
  \bibfield  {author} {\bibinfo {author} {\bibfnamefont {P.}~\bibnamefont
  {Fran{\c{c}}ois}}\ and\ \bibinfo {author} {\bibfnamefont {A.}~\bibnamefont
  {Zilman}},\ }\bibfield  {title} {\bibinfo {title} {Physical approaches to
  receptor sensing and ligand discrimination},\ }\href@noop {} {\bibfield
  {journal} {\bibinfo  {journal} {Current Opinion in Systems Biology}\ }\textbf
  {\bibinfo {volume} {18}},\ \bibinfo {pages} {111} (\bibinfo {year}
  {2019})}\BibitemShut {NoStop}%
\bibitem [{\citenamefont {Singh}\ and\ \citenamefont
  {Nemenman}(2015)}]{singh2015accurate}%
  \BibitemOpen
  \bibfield  {author} {\bibinfo {author} {\bibfnamefont {V.}~\bibnamefont
  {Singh}}\ and\ \bibinfo {author} {\bibfnamefont {I.}~\bibnamefont
  {Nemenman}},\ }\bibfield  {title} {\bibinfo {title} {Accurate sensing of
  multiple ligands with a single receptor},\ }\href@noop {} {\bibfield
  {journal} {\bibinfo  {journal} {arXiv preprint arXiv:1506.00288}\ } (\bibinfo
  {year} {2015})}\BibitemShut {NoStop}%
\bibitem [{\citenamefont {Singh}\ and\ \citenamefont
  {Nemenman}(2020)}]{singh2020universal}%
  \BibitemOpen
  \bibfield  {author} {\bibinfo {author} {\bibfnamefont {V.}~\bibnamefont
  {Singh}}\ and\ \bibinfo {author} {\bibfnamefont {I.}~\bibnamefont
  {Nemenman}},\ }\bibfield  {title} {\bibinfo {title} {Universal properties of
  concentration sensing in large ligand-receptor networks},\ }\href@noop {}
  {\bibfield  {journal} {\bibinfo  {journal} {Physical review letters}\
  }\textbf {\bibinfo {volume} {124}},\ \bibinfo {pages} {028101} (\bibinfo
  {year} {2020})}\BibitemShut {NoStop}%
\bibitem [{\citenamefont {Singh}\ and\ \citenamefont
  {Nemenman}(2017)}]{singh2017simple}%
  \BibitemOpen
  \bibfield  {author} {\bibinfo {author} {\bibfnamefont {V.}~\bibnamefont
  {Singh}}\ and\ \bibinfo {author} {\bibfnamefont {I.}~\bibnamefont
  {Nemenman}},\ }\bibfield  {title} {\bibinfo {title} {Simple biochemical
  networks allow accurate sensing of multiple ligands with a single receptor},\
  }\href@noop {} {\bibfield  {journal} {\bibinfo  {journal} {PLoS computational
  biology}\ }\textbf {\bibinfo {volume} {13}},\ \bibinfo {pages} {e1005490}
  (\bibinfo {year} {2017})}\BibitemShut {NoStop}%
\bibitem [{\citenamefont {Govern}\ and\ \citenamefont {ten
  Wolde}(2012)}]{govern2012fundamental}%
  \BibitemOpen
  \bibfield  {author} {\bibinfo {author} {\bibfnamefont {C.~C.}\ \bibnamefont
  {Govern}}\ and\ \bibinfo {author} {\bibfnamefont {P.~R.}\ \bibnamefont {ten
  Wolde}},\ }\bibfield  {title} {\bibinfo {title} {Fundamental limits on
  sensing chemical concentrations with linear biochemical networks},\
  }\href@noop {} {\bibfield  {journal} {\bibinfo  {journal} {Physical review
  letters}\ }\textbf {\bibinfo {volume} {109}},\ \bibinfo {pages} {218103}
  (\bibinfo {year} {2012})}\BibitemShut {NoStop}%
\bibitem [{\citenamefont {Ahuja}\ \emph {et~al.}(2020)\citenamefont {Ahuja},
  \citenamefont {Bhatnagar} \emph {et~al.}}]{ahuja2020capacity}%
  \BibitemOpen
  \bibfield  {author} {\bibinfo {author} {\bibfnamefont {M.}~\bibnamefont
  {Ahuja}}, \bibinfo {author} {\bibfnamefont {M.~R.}\ \bibnamefont
  {Bhatnagar}}, \emph {et~al.},\ }\bibfield  {title} {\bibinfo {title}
  {Capacity of ligand receptor channel with markovian symbol detection},\ }in\
  \href@noop {} {\emph {\bibinfo {booktitle} {2020 IEEE International
  Conference on Advanced Networks and Telecommunications Systems (ANTS)}}}\
  (\bibinfo {organization} {IEEE},\ \bibinfo {year} {2020})\ pp.\ \bibinfo
  {pages} {1--6}\BibitemShut {NoStop}%
\bibitem [{\citenamefont {Nguyen}\ \emph {et~al.}(2015)\citenamefont {Nguyen},
  \citenamefont {Dayan},\ and\ \citenamefont {Goodhill}}]{nguyen2015receptor}%
  \BibitemOpen
  \bibfield  {author} {\bibinfo {author} {\bibfnamefont {H.}~\bibnamefont
  {Nguyen}}, \bibinfo {author} {\bibfnamefont {P.}~\bibnamefont {Dayan}},\ and\
  \bibinfo {author} {\bibfnamefont {G.}~\bibnamefont {Goodhill}},\ }\bibfield
  {title} {\bibinfo {title} {How receptor diffusion influences gradient
  sensing},\ }\href@noop {} {\bibfield  {journal} {\bibinfo  {journal} {Journal
  of The Royal Society Interface}\ }\textbf {\bibinfo {volume} {12}},\ \bibinfo
  {pages} {20141097} (\bibinfo {year} {2015})}\BibitemShut {NoStop}%
\bibitem [{\citenamefont {Marder}(2011)}]{marder2011variability}%
  \BibitemOpen
  \bibfield  {author} {\bibinfo {author} {\bibfnamefont {E.}~\bibnamefont
  {Marder}},\ }\bibfield  {title} {\bibinfo {title} {Variability, compensation,
  and modulation in neurons and circuits},\ }\href@noop {} {\bibfield
  {journal} {\bibinfo  {journal} {Proceedings of the National Academy of
  Sciences}\ }\textbf {\bibinfo {volume} {108}},\ \bibinfo {pages} {15542}
  (\bibinfo {year} {2011})}\BibitemShut {NoStop}%
\bibitem [{\citenamefont {Hatakeyama}\ and\ \citenamefont
  {Kaneko}(2012)}]{hatakeyama2012generic}%
  \BibitemOpen
  \bibfield  {author} {\bibinfo {author} {\bibfnamefont {T.~S.}\ \bibnamefont
  {Hatakeyama}}\ and\ \bibinfo {author} {\bibfnamefont {K.}~\bibnamefont
  {Kaneko}},\ }\bibfield  {title} {\bibinfo {title} {Generic temperature
  compensation of biological clocks by autonomous regulation of catalyst
  concentration},\ }\href@noop {} {\bibfield  {journal} {\bibinfo  {journal}
  {Proceedings of the National Academy of Sciences}\ }\textbf {\bibinfo
  {volume} {109}},\ \bibinfo {pages} {8109} (\bibinfo {year}
  {2012})}\BibitemShut {NoStop}%
\bibitem [{\citenamefont {Kurosawa}\ and\ \citenamefont
  {Iwasa}(2005)}]{kurosawa2005temperature}%
  \BibitemOpen
  \bibfield  {author} {\bibinfo {author} {\bibfnamefont {G.}~\bibnamefont
  {Kurosawa}}\ and\ \bibinfo {author} {\bibfnamefont {Y.}~\bibnamefont
  {Iwasa}},\ }\bibfield  {title} {\bibinfo {title} {Temperature compensation in
  circadian clock models},\ }\href@noop {} {\bibfield  {journal} {\bibinfo
  {journal} {Journal of theoretical biology}\ }\textbf {\bibinfo {volume}
  {233}},\ \bibinfo {pages} {453} (\bibinfo {year} {2005})}\BibitemShut
  {NoStop}%
\bibitem [{\citenamefont {Thomas}\ and\ \citenamefont
  {Joy}(2006)}]{thomas2006elements}%
  \BibitemOpen
  \bibfield  {author} {\bibinfo {author} {\bibfnamefont {M.}~\bibnamefont
  {Thomas}}\ and\ \bibinfo {author} {\bibfnamefont {A.~T.}\ \bibnamefont
  {Joy}},\ }\href@noop {} {\emph {\bibinfo {title} {Elements of information
  theory}}}\ (\bibinfo  {publisher} {Wiley-Interscience},\ \bibinfo {year}
  {2006})\BibitemShut {NoStop}%
\bibitem [{\citenamefont {Minas}\ \emph {et~al.}(2020)\citenamefont {Minas},
  \citenamefont {Woodcock}, \citenamefont {Ashall}, \citenamefont {Harper},
  \citenamefont {White},\ and\ \citenamefont {Rand}}]{minas2020multiplexing}%
  \BibitemOpen
  \bibfield  {author} {\bibinfo {author} {\bibfnamefont {G.}~\bibnamefont
  {Minas}}, \bibinfo {author} {\bibfnamefont {D.~J.}\ \bibnamefont {Woodcock}},
  \bibinfo {author} {\bibfnamefont {L.}~\bibnamefont {Ashall}}, \bibinfo
  {author} {\bibfnamefont {C.~V.}\ \bibnamefont {Harper}}, \bibinfo {author}
  {\bibfnamefont {M.~R.}\ \bibnamefont {White}},\ and\ \bibinfo {author}
  {\bibfnamefont {D.~A.}\ \bibnamefont {Rand}},\ }\bibfield  {title} {\bibinfo
  {title} {Multiplexing information flow through dynamic signalling systems},\
  }\href@noop {} {\bibfield  {journal} {\bibinfo  {journal} {PLoS computational
  biology}\ }\textbf {\bibinfo {volume} {16}},\ \bibinfo {pages} {e1008076}
  (\bibinfo {year} {2020})}\BibitemShut {NoStop}%
\bibitem [{\citenamefont {Endres}\ and\ \citenamefont
  {Wingreen}(2009)}]{endres2009maximum}%
  \BibitemOpen
  \bibfield  {author} {\bibinfo {author} {\bibfnamefont {R.~G.}\ \bibnamefont
  {Endres}}\ and\ \bibinfo {author} {\bibfnamefont {N.~S.}\ \bibnamefont
  {Wingreen}},\ }\bibfield  {title} {\bibinfo {title} {Maximum likelihood and
  the single receptor},\ }\href@noop {} {\bibfield  {journal} {\bibinfo
  {journal} {Physical review letters}\ }\textbf {\bibinfo {volume} {103}},\
  \bibinfo {pages} {158101} (\bibinfo {year} {2009})}\BibitemShut {NoStop}%
\bibitem [{\citenamefont {Cover}\ and\ \citenamefont
  {Thomas}(2012)}]{Cover2012-zz}%
  \BibitemOpen
  \bibfield  {author} {\bibinfo {author} {\bibfnamefont {T.~M.}\ \bibnamefont
  {Cover}}\ and\ \bibinfo {author} {\bibfnamefont {J.~A.}\ \bibnamefont
  {Thomas}},\ }\href@noop {} {\emph {\bibinfo {title} {Elements of Information
  Theory}}}\ (\bibinfo  {publisher} {John Wiley \& Sons},\ \bibinfo {year}
  {2012})\BibitemShut {NoStop}%
\bibitem [{\citenamefont {Khamaru}\ and\ \citenamefont
  {Mazumder}(2019)}]{khamaru2019computation}%
  \BibitemOpen
  \bibfield  {author} {\bibinfo {author} {\bibfnamefont {K.}~\bibnamefont
  {Khamaru}}\ and\ \bibinfo {author} {\bibfnamefont {R.}~\bibnamefont
  {Mazumder}},\ }\bibfield  {title} {\bibinfo {title} {Computation of the
  maximum likelihood estimator in low-rank factor analysis},\ }\href@noop {}
  {\bibfield  {journal} {\bibinfo  {journal} {Mathematical Programming}\
  }\textbf {\bibinfo {volume} {176}},\ \bibinfo {pages} {279} (\bibinfo {year}
  {2019})}\BibitemShut {NoStop}%
\bibitem [{\citenamefont {Robertson}\ and\ \citenamefont
  {Symons}(2007)}]{robertson2007maximum}%
  \BibitemOpen
  \bibfield  {author} {\bibinfo {author} {\bibfnamefont {D.}~\bibnamefont
  {Robertson}}\ and\ \bibinfo {author} {\bibfnamefont {J.}~\bibnamefont
  {Symons}},\ }\bibfield  {title} {\bibinfo {title} {Maximum likelihood factor
  analysis with rank-deficient sample covariance matrices},\ }\href@noop {}
  {\bibfield  {journal} {\bibinfo  {journal} {Journal of Multivariate
  Analysis}\ }\textbf {\bibinfo {volume} {98}},\ \bibinfo {pages} {813}
  (\bibinfo {year} {2007})}\BibitemShut {NoStop}%
\bibitem [{\citenamefont {Weeks}\ \emph {et~al.}(1971)\citenamefont {Weeks},
  \citenamefont {Chandler},\ and\ \citenamefont
  {Andersen}}]{weeks1971perturbation}%
  \BibitemOpen
  \bibfield  {author} {\bibinfo {author} {\bibfnamefont {J.~D.}\ \bibnamefont
  {Weeks}}, \bibinfo {author} {\bibfnamefont {D.}~\bibnamefont {Chandler}},\
  and\ \bibinfo {author} {\bibfnamefont {H.~C.}\ \bibnamefont {Andersen}},\
  }\bibfield  {title} {\bibinfo {title} {Perturbation theory of the
  thermodynamic properties of simple liquids},\ }\href@noop {} {\bibfield
  {journal} {\bibinfo  {journal} {The Journal of Chemical Physics}\ }\textbf
  {\bibinfo {volume} {55}},\ \bibinfo {pages} {5422} (\bibinfo {year}
  {1971})}\BibitemShut {NoStop}%
\bibitem [{\citenamefont {Hopfield}(1974)}]{hopfield1974kinetic}%
  \BibitemOpen
  \bibfield  {author} {\bibinfo {author} {\bibfnamefont {J.~J.}\ \bibnamefont
  {Hopfield}},\ }\bibfield  {title} {\bibinfo {title} {Kinetic proofreading: a
  new mechanism for reducing errors in biosynthetic processes requiring high
  specificity},\ }\href@noop {} {\bibfield  {journal} {\bibinfo  {journal}
  {Proceedings of the National Academy of Sciences}\ }\textbf {\bibinfo
  {volume} {71}},\ \bibinfo {pages} {4135} (\bibinfo {year}
  {1974})}\BibitemShut {NoStop}%
\bibitem [{\citenamefont {Qian}(2006)}]{qian2006reducing}%
  \BibitemOpen
  \bibfield  {author} {\bibinfo {author} {\bibfnamefont {H.}~\bibnamefont
  {Qian}},\ }\bibfield  {title} {\bibinfo {title} {Reducing intrinsic
  biochemical noise in cells and its thermodynamic limit},\ }\href@noop {}
  {\bibfield  {journal} {\bibinfo  {journal} {Journal of molecular biology}\
  }\textbf {\bibinfo {volume} {362}},\ \bibinfo {pages} {387} (\bibinfo {year}
  {2006})}\BibitemShut {NoStop}%
\bibitem [{\citenamefont {Murugan}\ \emph {et~al.}(2012)\citenamefont
  {Murugan}, \citenamefont {Huse},\ and\ \citenamefont
  {Leibler}}]{murugan2012speed}%
  \BibitemOpen
  \bibfield  {author} {\bibinfo {author} {\bibfnamefont {A.}~\bibnamefont
  {Murugan}}, \bibinfo {author} {\bibfnamefont {D.~A.}\ \bibnamefont {Huse}},\
  and\ \bibinfo {author} {\bibfnamefont {S.}~\bibnamefont {Leibler}},\
  }\bibfield  {title} {\bibinfo {title} {Speed, dissipation, and error in
  kinetic proofreading},\ }\href@noop {} {\bibfield  {journal} {\bibinfo
  {journal} {Proceedings of the National Academy of Sciences}\ }\textbf
  {\bibinfo {volume} {109}},\ \bibinfo {pages} {12034} (\bibinfo {year}
  {2012})}\BibitemShut {NoStop}%
\bibitem [{\citenamefont {Horowitz}\ and\ \citenamefont
  {Sandberg}(2014)}]{horowitz2014second}%
  \BibitemOpen
  \bibfield  {author} {\bibinfo {author} {\bibfnamefont {J.~M.}\ \bibnamefont
  {Horowitz}}\ and\ \bibinfo {author} {\bibfnamefont {H.}~\bibnamefont
  {Sandberg}},\ }\bibfield  {title} {\bibinfo {title} {Second-law-like
  inequalities with information and their interpretations},\ }\href@noop {}
  {\bibfield  {journal} {\bibinfo  {journal} {New Journal of Physics}\ }\textbf
  {\bibinfo {volume} {16}},\ \bibinfo {pages} {125007} (\bibinfo {year}
  {2014})}\BibitemShut {NoStop}%
\bibitem [{Note1()}]{Note1}%
  \BibitemOpen
  \bibinfo {note} {Notice that this argument is general and the time-lag
  between $s(i)$ and $s(i+1)$ does not need to be equal to lag between $s(j)$
  and $s(j+1)$, for $1\leq i,j \leq n_t-1$. The effective time lags and
  effective time points $n_t$ are limited by the self-correlation of the
  trajectory $s(t)$ and by the property of the downstream kernel.}\BibitemShut
  {Stop}%
\bibitem [{\citenamefont {Arndt}(2000)}]{arndt2000yang}%
  \BibitemOpen
  \bibfield  {author} {\bibinfo {author} {\bibfnamefont {P.~F.}\ \bibnamefont
  {Arndt}},\ }\bibfield  {title} {\bibinfo {title} {Yang-lee theory for a
  nonequilibrium phase transition},\ }\href@noop {} {\bibfield  {journal}
  {\bibinfo  {journal} {Physical Review Letters}\ }\textbf {\bibinfo {volume}
  {84}},\ \bibinfo {pages} {814} (\bibinfo {year} {2000})}\BibitemShut
  {NoStop}%
\bibitem [{Note2()}]{Note2}%
  \BibitemOpen
  \bibinfo {note} {Derivatives in the ${\protect \bm {{\theta }}}$-space are
  evaluated as finite differentials with increments $\delta \theta =
  (4.8459\%\mu ^*, 5\%T^*, 5\%v_x^*)$. The 2-point trajectory probabilities
  used in the analysis are obtained by chopping a long simulation $10^{10}$
  steps (each step increases time by $dt=0.01$). Each 2-point trajectory is
  sampled from two time points with the time difference of $100$ steps which is
  equivalent to time lag $100 dt=1$.}\BibitemShut {Stop}%
\bibitem [{Note3()}]{Note3}%
  \BibitemOpen
  \bibinfo {note} {Notice that although the dynamics of the 6-state composite
  system is Markovian, its projection to the sensor's state space is no longer
  Markovian and have a history dependence.}\BibitemShut {Stop}%
\end{thebibliography}%

\end{document}